\documentclass[endfloats*,floatfix,a4paper,aip,jcp,nolongbibliography,preprint]{revtex4-2}
\usepackage[utf8]{inputenc}
\usepackage{amsfonts}
\usepackage{amssymb}
\usepackage{amsmath}
\usepackage{graphicx}
\usepackage[graphicx]{realboxes}
\usepackage{subcaption}
\usepackage{xcolor}
\usepackage[width=1\linewidth]{caption}
\usepackage{dcolumn}
\newcolumntype{d}[1]{D{.}{.}{#1}}

\newcommand{\mcc}[1]{\multicolumn{1}{c}{#1}}

\begin{document}

\title{Three-body potential and third virial coefficients for helium including relativistic and nuclear-motion effects}

\author{J. Lang}
\affiliation{Faculty of Chemistry, University of Warsaw, Pasteura 1, 02-093 Warsaw, Poland}
\email[Corresponding author: ]{jakub.lang@chem.uw.edu.pl}
\author{G. Garberoglio}
\affiliation{European Centre for Theoretical Studies in Nuclear Physics and Related Areas (FBK-ECT*), strada delle Tabarelle 286, I-38123, Trento, Italy.}
\affiliation{Trento Institute for Fundamental Physics and Applications (INFN-TIFPA), via Sommarive 14, I-38123, Trento, Italy.}
\author{M. Przybytek}
\affiliation{Faculty of Chemistry, University of Warsaw, Pasteura 1, 02-093 Warsaw, Poland}
\author{M. Jeziorska}
\affiliation{Faculty of Chemistry, University of Warsaw, Pasteura 1, 02-093 Warsaw, Poland}
\author{B. Jeziorski}
\affiliation{Faculty of Chemistry, University of Warsaw, Pasteura 1, 02-093 Warsaw, Poland}

\begin{abstract}
The non-additive three-body interaction potential for helium was computed using the coupled-cluster theory and the full configuration interaction method.
The obtained potential comprises an improved nonrelativistic Born--Oppenheimer energy and the leading relativistic and nuclear-motion corrections.
The mean absolute uncertainty of our calculations due to the incompleteness of the orbital basis set was determined employing complete-basis-set extrapolation techniques and was found to be 1.2\%.
For three helium atoms forming an equilateral triangle with the side length of 5.6~bohr -- a geometry close to the minimum of the total potential energy surface -- our three-body potential amounts to $-$90.6~mK, with an estimated uncertainty of 0.5~mK. 
An analytic function, developed to accurately fit the computed three-body interaction energies, was chosen to correctly describe the asymptotic behavior of the three-body potential for trimer configurations corresponding to both the three-atomic and the atom-diatom fragmentation channels.
For large triangles with sides $r_{12}$, $r_{23}$, and $r_{31}$, the potential takes correctly into account all angular terms decaying as $r_{12}^{-l} r_{23}^{-m} r_{31}^{-n}$ with \mbox{$l+m+n \le 14$} for the nonrelativistic Born--Oppenheimer energy and \mbox{$l+m+n \le 9$} for the post-Born--Oppenheimer corrections.
We also developed a short-range analytic function describing the local behavior of the total uncertainty of the computed three-body interaction energies. 
Using both fits we calculated the third pressure and acoustic virial coefficients for helium and their uncertainties for a wide range of temperatures. 
The results of these calculations were compared with available experimental data and with previous theoretical determinations. 
The estimated uncertainties of present calculations are 3-5 times smaller than those reported in the best previous work.
\end{abstract}

\maketitle

\section{Introduction}
\label{sec:introduction}

The recent redefinition of the SI units of energy (joule), temperature (kelvin), and amount of substance (mol) will enable more precise measurements of pressure and other thermophysical properties which are important in many areas of human activity, from vacuum manufacturing of semiconductor chips to aviation and marine navigation. 
Noble gases are systems particularly suitable for developing more accurate and practically convenient primary standards of pressure \cite{2020NatPh..16..177G} and temperature. \cite{gaiser2015dielectric}
For these gases, the major source of uncertainty is their non-ideal behavior that can be expressed through the well-understood virial expansion
\begin{equation}
\frac{p}{\rho R T} = 1 + B(T)\rho + C(T)\rho^2 + \cdots,
\end{equation}
where $p$ is the pressure, $T$ is the temperature, $\rho$ is the molar density, and $R$ is the gas constant, $R = N_A k_B$, with $N_A$ and $k_B $ being the Avogadro number and the Boltzmann constant, presently fixed at $6.022\,140\,76 \cdot 10^{23}$ mol$^{-1}$ and $1.380\,649 \cdot 10^{-23}$ J~K$^{-1}$, respectively. \cite{RevModPhys.93.025010}
The temperature-dependent functions, $B(T)$ and $C(T)$, are the second and the third virial coefficient, respectively, which can be computed from the pair and three-body potentials available from \emph{ab initio} quantum mechanical calculations. 
The second virial coefficient is usually enough to qualitatively describe the non-ideal behavior.
However, as the density increases, the contribution from the third virial coefficient becomes significant. 

Having only two electrons, helium is the simplest of the noble gases so \emph{ab initio} calculations with controlled accuracy are possible for its pair and three-body potential.
Aziz \emph{et al.} \cite{aziz1995ab} were the first to propose the use of \emph{ab initio} calculations in developing standards for measurements of thermophysical properties. 
Somewhat later, Hurly and Moldover \cite{hurly2000ab} calculated helium’s second virial coefficient $B(T)$, dilute-gas viscosity, and dilute-gas thermal conductivity with uncertainties smaller than those of the best experiments.
Their results were further improved by Hurly and Mehl. \cite{hurly20074he}
Bich \emph{et al.} \cite{bich2007ab} performed similar calculations of pair-potential-dependent quantities but considered also the third virial coefficient $C(T)$ using the Axilrod--Teller--Muto (ATM) \cite{Axilrod:43,Muto:43} representation of the three-body interaction energy. 
Nowadays, classical calculations of the third virial coefficient can be routinely done and are sufficiently accurate for high temperatures. \cite{garberoglio2009firstprincip}
However, when the temperature decreases, the classical approximation for $C(T)$ starts to break down and quantum effects become significant even for the room temperature, especially for a light system such as helium. \cite{garberoglio2009firstprincip,garberoglio2011improved,ram1973quantum,shaul2012semiclassical}
Classical results can be corrected down to some temperature limit by semiclassical methods, which can be based on the Wigner--Kirkwood expansion in powers of the Planck constant, \cite{kihara1953virial,dewitt1962analytic} or on the Feynman--Hibbs effective potentials. \cite{feynman1965pathintegrals,feynman1972statistical,guillot1998quantum}
For temperatures less than 50~K the semiclassical methods are not accurate enough and a full quantum mechanical approach is required. \cite{garberoglio2009firstprincip,shaul2012semiclassical}
While the quantum mechanical theory for the second virial coefficient has been known for a long time, \cite{hirschfelder1954molecular} the first fully quantum calculations of the third virial coefficient were performed only in the late sixties by Fosdick and Jordan  \cite{jordan1968three,fosdick1968monte} who employed a path-integral expression for $C(T)$ and a simple Lennard-Jones model for the helium pair potential. 
More recently, Harvey and one of the present authors (G.G.) further developed the path-integral method and calculated third virial coefficients using modern two- and three-body potentials. \cite{garberoglio2009firstprincip,garberoglio2011improved}
Calculations of higher-order virial coefficients of helium were also reported but all of them use some approximate form of the many-body interaction potentials beyond the three-body potential. \cite{shaul2012semiclassical,shaul2012path}

The theoretical knowledge of the interaction between two helium atoms has progressed so far that the two-body nonrelativistic Born--Oppenheimer (BO) potential is known with submillikelvin accuracy. \cite{Przybytek:17,czachorowski2020second} 
Various corrections beyond the BO approximation are known as well. \cite{Cencek:12,czachorowski2020second}
The relativistic, quantum-electrodynamic, adiabatic, and nonadiabatic corrections are currently available with relative error of less than 2\% in the well region. \cite{czachorowski2020second}
In the case of the three-body interaction the situation is different. 
The first three-body potential for helium was developed by Bruch and McGee in 1973. \cite{bruch1973calculations}
Already in this century, Lotrich and Szalewicz \cite{lotrich2000perturbation} published the three-body potential obtained using the symmetry-adapted perturbation theory (SAPT). \cite{jeziorski1994perturbation,szalewicz2022physical}
Somewhat later, the coupled-cluster theory with single, double, and noniterative triple excitations [CCSD(T)] \cite{nonitT1,nonitT2} was applied to the helium trimer by one of us (M.J.) and collaborators from the Szalewicz group. \cite{cencek2007three}
This work neglected the effects of excitations beyond the CCSD(T) level. 
In 2009, these effects where included in the work by Cencek \emph{et al.} \cite{Cencek:09}
These authors included contributions of higher excitations using the full configuration interaction (FCI) method \cite{Sherrill1999} with a small basis set [4s3p1d] and estimated the global error of their three-body potential at 2\%. \cite{Cencek:09}
So far all calculations of the three-body interactions have neglected the effect of post-BO corrections. 

In the present work, we shall present an analytic fit of an improved three-body BO interaction potential, and similar fits of the relativistic and adiabatic corrections to this potential obtained using the coupled-cluster theory and the FCI method. 
We shall provide also an analytic fit of the local uncertainty of the developed potential.
Finally, we shall report the third pressure and acoustic virial coefficients calculated   
using the three-body potentials obtained by us. 
 
Unless otherwise stated atomic units are used throughout the article.
The following 2018 CODATA values \cite{RevModPhys.93.025010} of fundamental constants and conversion factors are adopted: $1/137.035\,999\,084$ for the fine-structure constant $\alpha$, $7294.299\,541\,42\,m_e$ for the mass of the helium nucleus, $5.291\,772\,109\,03\times10^{-11}$~m for Bohr radius, and $315\,775.025$~K/$E_\mathrm{h}$ for the hartree to kelvin conversion factor.

\section{Calculation of three-body potentials}
\label{sec:calc3b}

\subsection{Definition of the potentials}

Let us consider a system comprising three identical atoms located at fixed positions in space specified by vectors $\mathbf{r}_I$, $I=1,2,3$.
The energy of the whole three-atomic system and the energies of subsystems composed of pairs of atoms at $\mathbf{r}_I$ and $\mathbf{r}_J$ depend on these atomic positions. 
This geometry dependence will be denoted by $E_3(\mathbf{r}_1,\mathbf{r}_2,\mathbf{r}_3)$ and $E_2(\mathbf{r}_I,\mathbf{r}_J)$, respectively.
Using the so-called many-body expansion, the energy of the whole system can be represented as
\begin{equation}\label{eq:entot}
E_3(\mathbf{r}_1,\mathbf{r}_2,\mathbf{r}_3)=
3E_1
+U_2(\mathbf{r}_1,\mathbf{r}_2)+U_2(\mathbf{r}_2,\mathbf{r}_3)+U_2(\mathbf{r}_3,\mathbf{r}_1)
+U_3(\mathbf{r}_1,\mathbf{r}_2,\mathbf{r}_3),
\end{equation}
where $E_1$ is the energy of a single atom, $U_2(\mathbf{r}_I,\mathbf{r}_J)$ is the pair potential defined by
\begin{equation}\label{eq:en2}
U_2(\mathbf{r}_I,\mathbf{r}_J)=E_2(\mathbf{r}_I,\mathbf{r}_J)-2E_1,
\end{equation}
and $U_3(\mathbf{r}_1,\mathbf{r}_2,\mathbf{r}_3)$ is the three-body potential that
represents a pair-wise non-additive part of the total interaction energy, $E_3(\mathbf{r}_1,\mathbf{r}_2,\mathbf{r}_3)-3E_1$.
Combining Eqs.~\eqref{eq:entot} and \eqref{eq:en2}, one obtains the formula
\begin{equation}\label{eq:en3}
U_3(\mathbf{r}_1,\mathbf{r_2},\mathbf{r}_3)=
 E_3(\mathbf{r}_1,\mathbf{r_2},\mathbf{r}_3)
-E_2(\mathbf{r}_1,\mathbf{r}_2)-E_2(\mathbf{r}_2,\mathbf{r}_3)-E_2(\mathbf{r}_3,\mathbf{r}_1)
+3E_1,
\end{equation}
which is convenient for direct calculation of the three-body potential using the supermolecular approach. 
As the energy of any atomic or molecular system in the absence of external fields is a property invariant to global translations and rotations, $U_3$ may be alternatively viewed as a function of a set of internal coordinates. 
For a three-atomic system, it is natural to choose the interatomic distances $r_{12}$, $r_{23}$, and $r_{31}$, where $r_{IJ}=|\mathbf{r}_J-\mathbf{r}_I|$, as the internal coordinates.
In the following text, the explicit dependence of $U_3$ on the interatomic distances will be often omitted for brevity.

Calculation of the BO component of the three-body potential, $U_3^\mathrm{BO}$, is straightforward using Eq.~\eqref{eq:en3} and identifying $E_3$, $E_2$, and $E_1$ with the BO energies calculated separately for the trimer, all atomic pairs, and single atoms, respectively. 
The supermolecular approach can be extended to obtain also the post-BO corrections to the potential by including the post-BO effects through the perturbation theory.
The energies $E_3$, $E_2$, and $E_1$ in Eq.~\eqref{eq:en3} are then identified with the perturbative corrections calculated separately for the trimer and its subsystems.
In this work we consider only the leading relativistic, $U_3^\mathrm{rel}$, and adiabatic, $U_3^\mathrm{ad}$, corrections to the three-body potential for helium.

For systems involving light atoms with small atomic numbers, the relativistic effects can be accounted for perturbatively by employing the approach based on the Breit-Pauli Hamiltonian. \cite{BeSal,Pachucki:04} 
This approach is accurate to the second order in the fine structure constant $\alpha$. 
In the case of the helium atom and small helium clusters, which are all closed-shell systems in their ground states, the spin-orbit and most of the spin-spin interactions vanish and the relativistic correction may be defined as the expectation value of the operator
\begin{equation}\label{rel:tot}
H^\mathrm{rel}
=H^\mathrm{mv}
+H^\mathrm{D1}
+H^\mathrm{D2}
+H^\mathrm{oo}
\end{equation}
calculated with the nonrelativistic wave function.
$H^\mathrm{mv}$ is the one-electron mass-velocity operator,
\begin{equation}\label{rel:mv}
H^\mathrm{mv} = 
-\frac{\alpha^2}{8} \sum_{i} \mathbf{p}_{i}^{4},
\end{equation}
$H^\mathrm{D1}$ is the one-electron Darwin operator,
\begin{equation}\label{rel:D1}
H^\mathrm{D1}
= \alpha^2 \frac{\pi}{2} \sum_{I}\sum_{i}\,Z_I\delta(\mathbf{r}_{Ii}),
\end{equation}
$H^\mathrm{D2}$ is the two-electron Darwin operator,
\begin{equation}\label{rel:D2}
H^\mathrm{D2}
= \alpha^2 \pi \sum_{i<j} \delta(\mathbf{r}_{ij}),
\end{equation}
and $H^\mathrm{oo}$ is the orbit-orbit operator,
\begin{equation}\label{rel:oo}
H^\mathrm{oo} 
= -\frac{\alpha^2}{2} \sum_{i<j}
\left( \frac{\mathbf{p}_i\cdot\mathbf{p}_j}{r_{ij}}
+ \frac{\mathbf{r}_{ij}\cdot(\mathbf{r}_{ij}\cdot\mathbf{p}_j)\,\mathbf{p}_i}{r_{ij}^3}
\right).
\end{equation}
In Eqs.~\eqref{rel:mv}--\eqref{rel:oo} indices $I$ and $i$ denote nuclei and electrons, respectively, $Z_I$ is the atomic number of a given nucleus, $\textbf{r}_{Ii}=\textbf{r}_i-\textbf{r}_I$ and $\textbf{r}_{ij}=\textbf{r}_{j}-\textbf{r}_{i}$ denote interparticle vectors, $\delta\left(\textbf{r}\right)$ is the Dirac delta function, and $\textbf{p}_i=-\mathrm{i}\nabla_{\mathbf{r}_i}$ is the momentum operator of the $i$-th electron. 
Expectation values of the one-electron relativistic operators, $H^\mathrm{mv}$ and $H^\mathrm{D1}$, are usually larger in magnitude than the expectation values of the two-electron relativistic operators, $H^\mathrm{D2}$ and $H^\mathrm{oo}$, have opposite sign, and cancel each other to a large extent. \cite{Piszczatowski2008} 
Therefore, it is advantageous to consider them always as a sum, introducing the Cowan--Griffin (CG) correction \cite{Cowan:76} defined as the expectation value of the operator
\begin{equation}\label{rel:CG}
H^\mathrm{CG}=H^\mathrm{mv}+H^\mathrm{D1}.
\end{equation}
In this work, we calculate separately three relativistic corrections to the three-body potential for helium, $U_3^\mathrm{CG}$, $U_3^\mathrm{D2}$, and $U_3^\mathrm{oo}$, resulting from the Cowan--Griffin, two-electron Darwin, and orbit-orbit operator, respectively. The final relativistic correction is then obtained as a sum of the components:
\begin{equation}\label{rel:components}
U_3^\mathrm{rel}=U_3^\mathrm{CG}+U_3^\mathrm{D2}+U_3^\mathrm{oo}.
\end{equation}

The adiabatic correction to the BO energy of a system [also known as the diagonal BO correction (DBOC)] can be calculated using the Born--Handy method \cite{Handy:86,Ioannou:96,Handy:96} as the expectation value of the nuclear kinetic energy operator
\begin{equation}\label{ad:tot}
H^\mathrm{ad}=\sum_I-\frac1{2m_I}\nabla_{\mathbf{r}_I}^2,
\end{equation}
where $m_I$ is the mass of nucleus $I$, and nuclear positions $\mathbf{r}_I$ are defined in the space-fixed coordinate frame. \cite{Kutzelnigg:97}
Calculation of the adiabatic correction involves differentiation of the electronic wave function with respect to nuclear coordinates and integration over electronic coordinates. 
Several approaches to carry out such calculation have been proposed, including numerical differentiation, \cite{Komasa:99,Valeev:03} variational calculation of the first-order wave function for the system perturbed by the nuclear potential, \cite{Pachucki:08,Przybytek:17} or using analytic wave function derivative techniques. \cite{Jensen:88,Gauss:06,Tajti:07,Tajti:09}
In this work, we employ the last approach.

\subsection{Computational details and extrapolation technique}

The BO energies were calculated using the coupled-cluster theory (CC) \cite{Bartlett2007} and the full configuration interaction (FCI) method. \cite{Sherrill1999}
The CC calculations were performed using the NCC module \cite{Matthews2015} from the CFOUR package, \cite{cfour,matthews2020coupled} which provides an implementation of a hierarchy of truncated CC methods, including CC with single and double excitations (CCSD), single, double, and triple excitations (CCSDT), and single, double, triple, and quadruple excitations (CCSDTQ), as well as variants of the truncated CC methods with noniterative treatment of the triple [CCSD(T)] \cite{nonitT1,nonitT2} and quadruple [CCSDT(Q)] \cite{nonitQ1,nonitQ2} excitations. 
The FCI calculations were performed using the Hector code \cite{przybytekFCI} interfaced with the Dalton 2.0 package \cite{daltonpaper,dalton2} for the Hartree--Fock orbitals and standard one- and two-electron integrals.

The post-BO corrections were calculated using only the CC theory. 
The calculations of the mass-velocity, and the one- and two-electron Darwin corrections were performed using the CFOUR program package up to the CCSDT level of theory. 
The orbit-orbit corrections, which are not available in the CFOUR package, were calculated using the Dalton 2018 program \cite{daltonpaper,dalton2018} as the first-order analytic derivatives of the CCSD(T) energy expression. \cite{Coriani:04}
The adiabatic corrections were calculated using the approach described in Ref.~\citenum{Gauss:06} and implemented in the CFOUR package. 
As the standalone version of the CFOUR code includes the adiabatic correction only at the CCSD level of CC theory, additional calculations at the CCSDT level were performed using the CFOUR code interfaced with the MRCC code \cite{kallay2020mrcc} as the CC solver.

The calculations of the BO energies and adiabatic corrections were performed using two families of correlation-consistent \cite{Dunning1989cc} polarized-valence Gaussian basis sets, which were specifically designed to accurately describe interaction energies in systems comprising ground-state helium atoms. \cite{Cencek:12,Przybytek:17}
In the following, these basis sets will be referred to as a$X$Z and d$X$Z, where the cardinal number $X$ is in the range $X=2,\dots,8$.
The letter 'a' or 'd' at the beginning of the name indicates whether the basis sets are singly or doubly augmented with diffuse functions with small exponents, which are necessary to correctly describe the wave function when the interatomic distances become large, that is, in the long-range limit.
The calculations of the relativistic corrections were performed using modified versions of the a$X$Z and d$X$Z basis sets. \cite{Cencek:12,Przybytek:17}
The modification consists in replacing the original set of $s$ functions by a common set of 22 uncontracted $s$ functions.
This modification is particularly important in the calculations of the one-electron relativistic corrections.
They are defined by singular operators, Eqs.~\eqref{rel:mv} and \eqref{rel:D1}, and require more flexible basis sets to properly describe the wave function in the vicinity of the nuclei.
The modified basis sets will be referred to as a$X$Zu and d$X$Zu, with the same range of the cardinal number $X$ as in the case of the original bases.

For any given atomic configuration, the values of the three-body BO potential and three-body post-BO corrections were calculated from Eq.~\eqref{eq:en3} using the energies or energy corrections, $E_3$, $E_2$, and $E_1$, computed at the same level of theory and with the same basis set.
In particular, the energies for the trimer, all atomic pairs, and single atoms were computed with the full basis set of the trimer, that is, the set comprising functions centered at positions of all three atoms in the system.
This approach, the so-called counterpoise method, is needed to remove the basis set superposition error (BSSE). \cite{Boys1970,Mierzwicki2003}
Note that, within this approach, the energy of a single helium atom is no longer constant and depends slightly on the position of the remaining atoms. 
Therefore, the atomic term $3E_1$ in Eq.~\eqref{eq:en3} must be replaced by the sum of three, in general different, terms coming from each atom. 
Similarly, the two-body energies $E_2(\mathbf{r}_I,\mathbf{r}_J)$ depend not only on the internuclear distance $r_{IJ}$, but also slightly on the position of the third atom. 
In practice, each calculation of the three-body energy requires seven different calculations using the same three-atomic basis set.  

To reduce the basis set incompleteness error (BSIE) in the results obtained with finite basis sets, and to assess the uncertainty of the \emph{ab initio} calculations, we employed the extrapolation technique to approximate the values of the potentials at the complete basis set (CBS) limit.
To this end, we assumed the following extrapolation formula,
\begin{equation}\label{eq:cbs}
U_3^\mathrm{Y}(X)=U_3^\mathrm{Y}(\infty)+A\,X^{-n^\mathrm{Y}},
\end{equation}
where $U_3^\mathrm{Y}(X)$, $\mathrm{Y}\in\{\mathrm{BO},\;\mathrm{CG},\;\mathrm{D2},\;\mathrm{oo},\;\mathrm{ad}\}$, is the value of a given three-body potential calculated using basis set with the cardinal number $X$, and $U_3^\mathrm{Y}(\infty)$ denotes the value of this potential at the CBS limit.
The rate of the CBS convergence, characterized by the value of the exponent $n^\mathrm{Y}$ in Eq.~\eqref{eq:cbs}, depends on the type of the potential.
In the case of the BO calculations, we adopted $n^\mathrm{BO}=3$, as recommended for the extrapolation of BO correlation energies. \cite{Halkier1998,Helgaker2008}
As was shown by Kutzelnigg \cite{Kutzelnigg08} and verified numerically several times, \cite{Salomonson89,Ottschofski97,Halkier00} the two-electron Darwin correction converges to the CBS limit very slowly, as $X^{-1}$. 
Following this observation we chose $n^\mathrm{D2}=1$. 
Unfortunately, there exists no theoretical justification for a proper choice of the exponent $n^\mathrm{Y}$ for other post-BO corrections. 
Przybytek \emph{et al.} \cite{Cencek:12,Przybytek:17} addressed this problem by investigating basis set convergence of the post-BO corrections in the case of the helium atom, for which very accurate reference values are known. \cite{DRAKE19887} 
They obtained $n^\mathrm{CG}=1$, $n^\mathrm{oo}=3/2$, and $n^\mathrm{ad}=3$. 
In this work, we adopted these values of exponents for use in the CBS extrapolations.
Knowing the value of the exponent $n^\mathrm{Y}$ and using results from two consecutive basis sets, with cardinal numbers $(X-1)$ and $X$, the value of a given potential at the CBS limit, $U_3^\mathrm{Y}(\infty)$, is easily obtained from the formula,
\begin{equation}\label{eq:cbstwopoint}
U_3^\mathrm{Y}(\infty)=U_3^\mathrm{Y}(X)+\frac{U_3^\mathrm{Y}(X)-U_3^\mathrm{Y}(X-1)}{\left(1-1/X\right)^{-n^\mathrm{Y}}-1}.
\end{equation}
In further text, by d$[(X-1)X]$Z we denote the values extrapolated from the results obtained from calculations with the d$(X-1)$Z and d$X$Z basis sets.
A similar notation is used in the case of the other families of basis sets used in this work.

To establish computationally efficient strategies capable of producing reliable and highly accurate results for the three-body energies, we conducted a series of tests for a selected set of configurations.
Test calculations were performed for four configurations where all atoms form an equilateral triangle with the side length equal to $R_\mathrm{side}=$ 4, 5.6, 7, or 9~bohr, and two centrosymmetric linear configurations with the distance between central atom and two outer atoms equal to $R_\mathrm{sep}=$ 4 or 5.6~bohr. 
These test configurations will be further referred to as Equilat$(R_\mathrm{side})$ and LinSym$(R_\mathrm{sep})$, respectively.
In the following, we discuss in detail results, shown in Tables~\ref{tab:bo-conv}, \ref{tab:rel-conv}, and \ref{tab:ad-conv}, for two configurations, namely Equilat(5.6) and LinSym(5.6), where at least two pairs of atoms are separated by the distance of 5.6~bohr, which is close to the minimum position of the pair potential for helium. 
Final recommended results for all test configurations are presented in Tables~\ref{tab:bo-final} and \ref{tab:post-final}.
The developed strategies were then employed to calculate the potentials for all configurations, generated as described in Sec.~\ref{sec:fit3b}, which were used to produce analytic fits.
A list of \emph{ab initio} values and estimated uncertainties of the potentials for these configurations is included in the ESI.$^\dag$

\subsection{BO energy}

\begin{table*}[t]
\small
\caption{\ Basis set convergence of contributions to the three-body BO potential for helium, defined in Eq.~\eqref{eq:bo}, for two test configurations.
Calculations were carried out using basis sets from the a$X$Z and d$X$Z families. 
Columns denoted as ``extr.''\ contain extrapolations performed using Eq.~\eqref{eq:cbstwopoint} applied to the results from the column to the left.
The energy unit is mK.}
\label{tab:bo-conv}
\begin{tabular*}{\textwidth}{@{\extracolsep{\fill}}cd{4.4}d{3.4}d{3.4}d{2.4}d{2.4}d{2.4}}
\hline \\[-0.7ex] 
basis set & 
\mcc{$U_3^\mathrm{BO}[\mathrm{HF}]$} &
\mcc{$\Delta U_3^\mathrm{BO}[\mathrm{CCSD(T)}]$} & \mcc{extr.} &
\mcc{$\Delta U_3^\mathrm{BO}[\mathrm{CCSDT}]$} & \mcc{extr.} &
\mcc{$\Delta U_3^\mathrm{BO}[\mathrm{FCI}]$} \\[1ex] 
\hline
\\[-1ex]
\multicolumn{7}{c}{Equilat(5.6)} \\
\\[-1ex]
a2Z &   -273.0230 &    120.9362 &   \mcc{$-$} &      7.8449 &   \mcc{$-$} &      1.1924 \\
a3Z &   -273.6202 &    168.8894 &    189.0802 &      8.3353 &      8.5417 &      1.2917 \\
a4Z &   -274.0328 &    173.1385 &    176.2392 &      7.9361 &      7.6449 &             \\
a5Z &   -274.3012 &    174.5307 &    175.9914 &      7.7894 &      7.6353 &             \\
a6Z &   -274.4074 &    175.1877 &    176.0900 &      7.6900 &      7.5535 &             \\
a7Z &   -274.4232 &    175.4892 &    176.0022 &             &             &             \\
a8Z &   -274.4496 &             &             &             &             &             \\[1em]
d2Z &   -274.8245 &    128.0188 &   \mcc{$-$} &      7.8197 &   \mcc{$-$} &      1.1111 \\
d3Z &   -274.3816 &    172.1405 &    190.7181 &      8.6678 &      9.0249 &      1.4301 \\
d4Z &   -274.4032 &    174.6446 &    176.4720 &      7.9084 &      7.3543 &             \\
d5Z &   -274.4829 &    174.7888 &    174.9400 &      7.6619 &      7.4033 &             \\
d6Z &   -274.4941 &    175.0866 &    175.4956 &      7.5775 &      7.4615 &             \\
d7Z &   -274.5025 &    175.2987 &    175.6594 &             &             &             \\
d8Z &   -274.5053 &             &             &             &             &             \\
\hline
\\[-1ex]
\multicolumn{7}{c}{LinSym(5.6)} \\
\\[-1ex]
a2Z &      1.5103 &    -15.1449 &   \mcc{$-$} &     -1.1703 &   \mcc{$-$} &     -0.3647 \\
a3Z &      1.4205 &    -16.8630 &    -17.5864 &     -1.6412 &     -1.8394 &     -0.4621 \\
a4Z &      1.5209 &    -17.5433 &    -18.0398 &     -1.7128 &     -1.7650 &             \\
a5Z &      1.5183 &    -17.6372 &    -17.7358 &     -1.7223 &     -1.7323 &             \\
a6Z &      1.5093 &    -17.7288 &    -17.8547 &     -1.7211 &     -1.7196 &             \\
a7Z &      1.5121 &    -17.7556 &    -17.8012 &             &             &             \\
a8Z &      1.5173 &             &             &             &             &             \\[1em]
d2Z &      1.8193 &    -14.5672 &   \mcc{$-$} &     -1.1835 &   \mcc{$-$} &     -0.3482 \\
d3Z &      1.4670 &    -17.1296 &    -18.2085 &     -1.6897 &     -1.9028 &     -0.4936 \\
d4Z &      1.5570 &    -17.7011 &    -18.1182 &     -1.7339 &     -1.7661 &             \\
d5Z &      1.5208 &    -17.8465 &    -17.9991 &     -1.7313 &     -1.7285 &             \\
d6Z &      1.5200 &    -17.8722 &    -17.9073 &     -1.7250 &     -1.7163 &             \\
d7Z &      1.5174 &    -17.8873 &    -17.9130 &             &             &             \\
d8Z &      1.5184 &             &             &             &             &             \\
\hline
\end{tabular*}
\end{table*}

In the calculations of the three-body BO potential we adopted a hybrid approach that relies upon splitting the calculated quantity into several contributions that are obtained using increasing levels of theory while employing basis sets of decreasing size.
We used a scheme comprising four steps
\begin{equation}\label{eq:bo}
U_3^\mathrm{BO}
= U_3^\mathrm{BO}[\mathrm{HF}] 
+ \Delta U_3^\mathrm{BO}[\mathrm{CCSD(T)}] 
+ \Delta U_3^\mathrm{BO}[\mathrm{CCSDT}] 
+ \Delta U_3^\mathrm{BO}[\mathrm{FCI}],
\end{equation}
where the last three terms are defined as
\begin{align}
\label{eq:bo1}
\Delta U_3^\mathrm{BO}[\mathrm{CCSD(T)}] &
= U_3^\mathrm{BO}[\mathrm{CCSD(T)}] 
- U_3^\mathrm{BO}[\mathrm{HF}],
\\
\label{eq:bo2}
\Delta U_3^\mathrm{BO}[\mathrm{CCSDT}] &
= U_3^\mathrm{BO}[\mathrm{CCSDT}] 
- U_3^\mathrm{BO}[\mathrm{CCSD(T)}],
\\
\label{eq:bo3}
\Delta U_3^\mathrm{BO}[\mathrm{FCI}] &
= U_3^\mathrm{BO}[\mathrm{FCI}] 
- U_3^\mathrm{BO}[\mathrm{CCSDT}].
\end{align}
In these equations $U_3^\mathrm{BO}[\mathrm{M}]$ represents the three-body BO potential calculated using method M, with M$\,=\,$HF denoting the Hartree--Fock method. 
It is understood that both values on the right hand sides of Eqs.~\eqref{eq:bo1}--\eqref{eq:bo3} are obtained with the same basis set, which in turn may be different for each contribution in Eq.~\eqref{eq:bo}.
We also considered alternative schemes with steps involving results obtained at CCSDT(Q) and CCSDTQ levels of theory.
After extensive testing, we found that these methods may be numerically unstable providing unreliable BO interaction energies for some atomic configurations.
Moreover, since the CCSDT(Q) and CCSDTQ calculations cannot be performed with basis sets significantly larger than the ones used in the FCI calculations, these alternative schemes are not competitive with the scheme defined in Eq.~\eqref{eq:bo}.

\begin{table*}
\scriptsize
\caption{\ Comparison of the recommended values of contributions to the three-body BO potential obtained in this work with the results of Ref.~\citenum{Cencek:09}.
Numbers in parentheses are uncertainties of the rightmost digits. 
The energy unit is mK.}
\label{tab:bo-final}
\begin{tabular*}{\textwidth}{@{\extracolsep{\fill}}ld{6.7}d{4.7}d{3.8}d{2.6}d{3.6}d{6.6}}
\hline\\[-0.7ex]
configuration &
\mcc{$U_3^\mathrm{BO}[\mathrm{HF}]$} &
\mcc{$\Delta U_3^\mathrm{BO}[\mathrm{CCSD(T)}]$} &
\mcc{$\Delta U_3^\mathrm{BO}[\mathrm{CCSDT}]$} &
\mcc{$\Delta U_3^\mathrm{BO}[\mathrm{FCI}]$} &
\mcc{ $\Delta U_3^\mathrm{BO}[\mathrm{CCSDT}]+\Delta U_3^\mathrm{BO}[\mathrm{FCI}]$}&
\mcc{$U_3^\mathrm{BO}$} 
\\[1ex]
\hline 
\\[-1ex]
Equilat(4)
& -61998.2(15) & 5392.(101) & 224.(7) & 14.5(25) & 238.(7) & -56368.(101) \\
Ref.~\citenum{Cencek:09} & -61990. & 5430.(40) &&& 283.(57)& -56277.(97) \\[1em]
Equilat(5.6)
& -274.494(11) & 175.5(4) & 7.40(26) & 1.29(10) & 8.7(3) & -90.3(5) \\
Ref.~\citenum{Cencek:09} & -274.3 & 175.8(2) &&& 10.0(13) & -88.3(17) \\[1em]
Equilat(7)
& -1.8508(1) & 16.811(24) & 0.911(18) & 0.22(3) & 1.13(4) & 16.09(4) \\
Ref.~\citenum{Cencek:09} & -1.85 & 16.72(9) &&& 1.19(7) & 16.06(17) \\[1em]
Equilat(9)
& -0.0012(1) & 1.6342(6) & 0.1021(11) & 0.025(4) & 0.127(4) & 1.760(4) \\[1em]
LinSym(4)
& 1654.0(7) & 181.(11) & -9.29(21) & -7.8(11) & -17.1(11) & 1818.(12) \\[1em]
LinSym(5.6)
& 1.5200(8) & -17.91(4) & -1.7285(28) & -0.46(10) & -2.19(10) & -18.58(10) \\
Ref.~\citenum{Cencek:09} & 1.52 & -18.04(6) &&& -2.12(7) & -18.59(18) \\[1ex]
\hline
\end{tabular*}
\end{table*}

Inspection of the basis set convergence pattern presented in Table~\ref{tab:bo-conv} can be helpful to show how the recommended values of the $U_3^\mathrm{BO}[\mathrm{HF}]$, $\Delta U_3^\mathrm{BO}[\mathrm{CCSD(T)}]$, $\Delta U_3^\mathrm{BO}[\mathrm{CCSDT}]$, and $\Delta U_3^\mathrm{BO}[\mathrm{FCI}]$ contributions were determined, and their uncertainty estimated.  
The Hartree--Fock contribution, $U_3^\mathrm{BO}[\mathrm{HF}]$, converges very fast with the increasing value of the basis set's cardinal number $X$.
In the case of the Equilat(5.6) configuration, the difference between values obtained with $X=6$ and $X=8$ for the a$X$Z family is 0.04~mK (0.015\% of the a8Z result), and for the d$X$Z family about 0.01~mK (0.004\% of the d8Z result). 
Therefore, we consider the $U_3^\mathrm{BO}[\mathrm{HF}]$ contribution calculated with the d6Z basis set as converged, and assign it the uncertainty equal to the absolute difference between the d5Z and d6Z results. 
This uncertainty is an order of magnitude smaller then the uncertainty of the correlation contribution and has practically no effect on the uncertainty of the total three-body potential.   

\begin{figure}[ht]
\centering
\includegraphics[width=1\columnwidth]{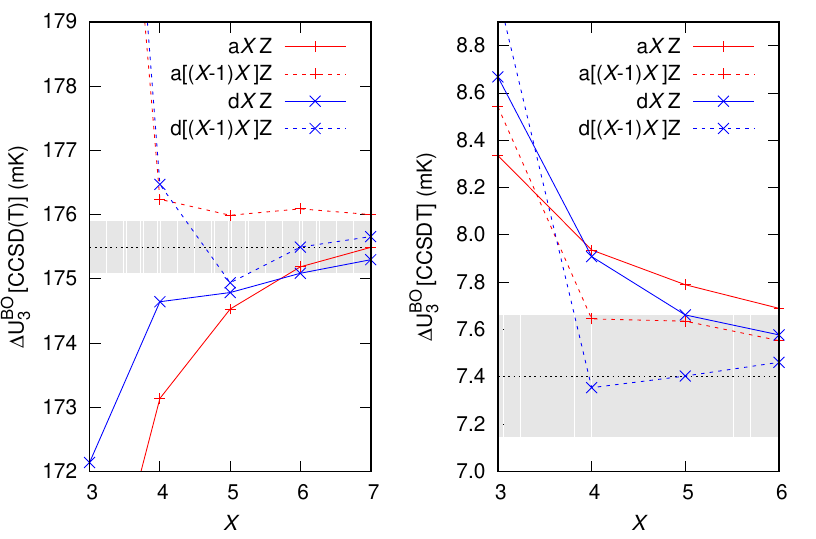}
\caption{
Basis set convergence of the $\Delta U_3^\mathrm{BO}[\mathrm{CCSD(T)}]$ (left panel) and $\Delta U_3^\mathrm{BO}[\mathrm{CCSDT}]$ (right panel) contributions to the three-body BO potential for helium calculated for the Equilat(5.6) configuration using basis sets from the a$X$Z and d$X$Z families.
Horizontal dotted line represents recommended value of the given correction, and the shaded area is the range of its estimated uncertainty.
}
\label{fig:bo-extr}
\end{figure}

To investigate the $\Delta U_3^\mathrm{BO}[\mathrm{CCSD(T)}]$ contribution, we performed CCSD(T) calculations using basis sets with cardinal numbers up to $X=7$.
A typical basis set convergence of $\Delta U_3^\mathrm{BO}[\mathrm{CCSD(T)}]$ obtained with the a$X$Z and d$X$Z families of basis sets, as well as the convergence of the corresponding a$[(X-1)X]$Z and d$[(X-1)X]$Z extrapolants, is illustrated in the left panel of Fig.~\ref{fig:bo-extr} using the Equilat(5.6) configuration as an example.
The convergence of the unextrapolated results is rather smooth and monotonic.
The values of the corresponding extrapolants seem to stabilize for $X\geq5$ and start to converge toward each other as the cardinal number $X$ increases.
For the majority of test configurations, the d(56)Z and d(67)Z extrapolants are closer to each other than the unextrapolated d6Z and d7Z values. 
Moreover, the d(56)Z extrapolant is always closer to the d(67)Z extrapolant than to the d6Z result.
In the case of the Equilat(5.6) configuration, for instance, the difference between the d(56)Z and d(67)Z extrapolants is 0.16~mK while the difference between the d(56)Z and d6Z results is 0.41~mK.
All these observations suggest that the extrapolation technique is able to provide reliable approximations to the CBS limit of the $\Delta U_3^\mathrm{BO}[\mathrm{CCSD(T)}]$ contribution.
In the end, we consider the extrapolated d(56)Z results as converged with the uncertainty estimated as the absolute difference between the d6Z result and the d(56)Z extrapolant. 

Cencek \emph{et al.} \cite{Cencek:09} pointed out that the basis set convergence of the CCSDT energies has not been explicitly investigated in the literature and it may be different, and most probably slower, than the convergence of the CCSD(T) energies. 
Therefore, they recommended performing two-point extrapolations of $\Delta U_3[\mathrm{CCSDT}]$, defined as in Eq.~\eqref{eq:cbstwopoint}, using exponents $n^\mathrm{BO}=2$ and $n^\mathrm{BO}=3$, and then selecting the result that leads to a larger error estimation.
A reason for this recommendation could have been the use of the 
standard Dunning's aug-cc-pV$X$Z basis sets for helium. \cite{Dunning1989cc,Dunning1994he}
These basis sets were constructed to achieve accurate predictions of interactions in a system containing helium atoms through an accurate description of electrical properties of a single atom.
Such design principle may be insufficient to properly account for small, higher-level correlation effects in the weak non-additive three-body interaction energies for helium.
By contrast, our a$X$Z and d$X$Z basis sets were optimized specifically for the interaction energies. \cite{Cencek:12}
The basis set convergence pattern of the $\Delta U_3^\mathrm{BO}[\mathrm{CCSDT]}$ results and the dependence on the cardinal number $X$ of the extrapolants obtained by us assuming $n^\mathrm{BO}=3$ is qualitatively similar to that of the $\Delta U_3^\mathrm{BO}[\mathrm{CCSD(T)}]$ contribution.
As can be seen in the right panel of Fig.~\ref{fig:bo-extr}, the apparent stabilization of extrapolants for $\Delta U_3^\mathrm{BO}[\mathrm{CCSDT}]$ in the case of the Equilat(5.6) configuration is even more pronounced than for $\Delta U_3^\mathrm{BO}[\mathrm{CCSD(T)}]$ and occurs earlier, for $X\geq4$.
A similar behavior is observed also for the remaining test configurations.
Therefore, we assume that the two-point extrapolation technique with $n^\mathrm{BO}=3$ is applicable also in the case of the $\Delta U_3^\mathrm{BO}[\mathrm{CCSDT}]$ contribution.
As the CCSDT calculations are computationally more expensive and the apparent stabilization of extrapolants occurs earlier, we use smaller basis sets in approximating the CBS limit of $\Delta U_3^\mathrm{BO}[\mathrm{CCSDT}]$.
We consider the extrapolated d(45)Z results as converged with the uncertainty estimated as the absolute difference between the d5Z value and the d(45)Z extrapolant. 
Note, that in contrast to the previous calculations of Cencek \emph{et al.}, \cite{Cencek:09} who calculated $\Delta U_3^\mathrm{BO}[\mathrm{CCSDT}]$ only for a few selected configurations, we calculated this correction for all configurations used for the final fitting of the three-body BO potential. 

The last contribution to the BO potential is the FCI correction $\Delta U_3^\mathrm{BO}[\mathrm{FCI}]$.
Due to numerical complexity of the FCI calculations for a six-electron system, we were able to obtain this correction using basis sets with the cardinal numbers only up to $X=3$.
Moreover, since basis sets with $X=2$ and $X=3$ are too small to ensure reliable extrapolations, we consider the a3Z results as our recommended values and estimate the uncertainty as the absolute difference between the a2Z and a3Z results.
Note, that our a3Z basis set with the composition [5s3p2d] is still larger than the [4s3p1d] basis set employed in the FCI calculations in Ref.~\citenum{Cencek:09}.

In Table~\ref{tab:bo-final}, we collect our recommended values of the contributions to the three-body BO potential, together with their estimated uncertainties, calculated for all test configurations.
In each case, the final value of $U_3^\mathrm{BO}$ is obtained by summing all the contributions as defined in Eq.~\eqref{eq:bo}, and its uncertainty, $\sigma_3^\mathrm{BO}$, is calculated as the sum in quadrature of the uncertainties of each contribution.
The value of the three-body BO potential is dominated by the sum of the HF energy and the CCSD(T) correlation energy.
The effect of iterative, post-CCSD(T) triples, $\Delta U_3^\mathrm{BO}[\mathrm{CCSDT}]$, corresponds to at most 10\% of $U_3^\mathrm{BO}$, which is similar to the estimation given in Ref.~\citenum{Cencek:09}, and the contribution due to higher correlation effects is smaller than 3\%.
Overall, the importance of the total post-CCSD(T) contribution, $\Delta U_3^\mathrm{BO}[\mathrm{CCSDT}]+\Delta U_3^\mathrm{BO}[\mathrm{FCI}]$, increases with the size of the considered triangle of atoms. 
The estimated uncertainty of the total BO component of the potential, $\sigma_3^\mathrm{BO}$, is usually dominated by the uncertainty of $\Delta U_3^\mathrm{BO}[\mathrm{CCSD(T)}]$ but for configurations with large interatomic distances $\Delta U_3^\mathrm{BO}[\mathrm{FCI}]$ becomes also important.

In Table~\ref{tab:bo-final}, we also compare our predictions with the results reported by Cencek \emph{et al.} \cite{Cencek:09}
Our benchmark values lie very close to the results from Ref.~\citenum{Cencek:09}, with the differences always being within the range of combined error bounds.
However, with the exception of the configuration Equilat(4), our estimated uncertainties are 2-4 times smaller than those published in Ref.~\citenum{Cencek:09}.

\begin{table}[t!]
\small
\caption{\ Basis set convergence of the components of the three-body relativistic correction for helium, defined in Eq.~\eqref{rel:components}, for two test configurations.
Only calculations carried out using basis sets from the d$X$Zu family are presented. 
Column denoted as ``extr.''\ contain extrapolations performed using Eq.~\eqref{eq:cbstwopoint} applied to the results from the column to the left.
Entries in the last column are defined by the formula $\delta[\mathrm{CCSDT}]=(U_3^\mathrm{Y}[\mathrm{CCSDT}]-U_3^\mathrm{Y}[\mathrm{CCSD(T)}])/U_3^\mathrm{Y}[\mathrm{CCSD(T)}]\times100\%$.
The energy unit is mK.}
\label{tab:rel-conv}
\begin{tabular*}{0.48\textwidth}{@{\extracolsep{\fill}}cd{2.5}d{2.5}d{2.3}}
\hline \\[-0.7ex] 
basis set &
\mcc{$U_3^\mathrm{Y}[\mathrm{CCSD(T)}]$} & \mcc{extr.} &
\mcc{$\delta[\mathrm{CCSDT}]$} \\[1ex]
\hline
\\[-1ex]
\multicolumn{4}{c}{Equilat(5.6)}\\
\\[-1ex]
\multicolumn{2}{c}{$\mathrm{Y}=\mathrm{CG}$} \\  
d2Zu &     0.04287 &   \mcc{$-$} &     2.4 \,\% \\
d3Zu &     0.03560 &     0.02105 &     1.8 \,\% \\
d4Zu &     0.03664 &     0.03975 &     1.6 \,\% \\
d5Zu &     0.03742 &     0.04055 &     1.5 \,\% \\
d6Zu &     0.03786 &     0.04007 &              \\
\\                                           
\multicolumn{2}{c}{$\mathrm{Y}=\mathrm{D2}$} \\
d2Zu &    -0.05686 &   \mcc{$-$} &    -4.9 \,\% \\
d3Zu &    -0.04319 &    -0.01585 &    -6.5 \,\% \\
d4Zu &    -0.04025 &    -0.03143 &    -6.0 \,\% \\
d5Zu &    -0.03887 &    -0.03335 &    -5.8 \,\% \\
d6Zu &    -0.03796 &    -0.03339 &              \\
\\                                        
\multicolumn{2}{c}{$\mathrm{Y}=\mathrm{oo}$} \\      
d2Zu &    -0.05929 &   \mcc{$-$} & \\
d3Zu &    -0.07079 &    -0.08454 & \\
d4Zu &    -0.06835 &    -0.06383 & \\
d5Zu &    -0.06751 &    -0.06540 & \\
d6Zu &    -0.06757 &    -0.06774 & \\
\hline
\\[-1ex]
\multicolumn{4}{c}{LinSym(5.6)}\\
\\[-1ex]
\multicolumn{2}{c}{$\mathrm{Y}=\mathrm{CG}$} \\    
d2Zu &     0.00334 &   \mcc{$-$} &    -2.8 \,\% \\
d3Zu &     0.00434 &     0.00632 &     1.2 \,\% \\
d4Zu &     0.00447 &     0.00488 &     2.4 \,\% \\
d5Zu &     0.00453 &     0.00479 &     1.9 \,\% \\
d6Zu &     0.00457 &     0.00476 &              \\
\\                                           
\multicolumn{2}{c}{$\mathrm{Y}=\mathrm{D2}$} \\
d2Zu &    -0.00065 &   \mcc{$-$} &    53.1 \,\% \\
d3Zu &    -0.00126 &    -0.00248 &    33.1 \,\% \\
d4Zu &    -0.00128 &    -0.00132 &    31.9 \,\% \\
d5Zu &    -0.00129 &    -0.00134 &    29.7 \,\% \\
d6Zu &    -0.00128 &    -0.00127 &              \\
\\                                        
\multicolumn{2}{c}{$\mathrm{Y}=\mathrm{oo}$} \\      
d2Zu &     0.06320 &   \mcc{$-$} & \\
d3Zu &     0.06263 &     0.06195 & \\
d4Zu &     0.06404 &     0.06664 & \\
d5Zu &     0.06398 &     0.06383 & \\
d6Zu &     0.06555 &     0.07055 & \\
\hline
\end{tabular*}
\end{table}

\begin{table}[t]
\small
\caption{\ Basis set convergence of the three-body adiabatic correction for helium for two test configurations.
Only calculations carried out using basis sets from the d$X$Z family are presented. 
Column denoted as ``extr.''\ contain extrapolations performed using Eq.~\eqref{eq:cbstwopoint} applied to the results from the column to the left.
Entries in the last column are defined by the formula $\delta[\mathrm{CCSDT}]=(U_3^\mathrm{ad}[\mathrm{CCSDT}]-U_3^\mathrm{ad}[\mathrm{CCSD}])/U_3^\mathrm{ad}[\mathrm{CCSD}]\times100\%$.
The energy unit is mK.}
\label{tab:ad-conv}
\begin{tabular*}{0.48\textwidth}{@{\extracolsep{\fill}}cd{2.5}d{2.5}d{2.3}} 
\hline \\[-0.7ex]
basis set &
\mcc{$U_3^\mathrm{ad}[\mathrm{CCSD}]$} & \mcc{extr.} & 
\mcc{$\delta[\mathrm{CCSDT}]$} \\[1ex]
\hline
\\[-1ex]
\multicolumn{4}{c}{Equilat(5.6)}\\
\\[-1ex]
d2Z &    -0.28478 &   \mcc{$-$} &     24.7 \,\% \\
d3Z &    -0.24832 &    -0.23298 &     28.6 \,\% \\
d4Z &    -0.24970 &    -0.25070 &               \\
d5Z &    -0.25029 &    -0.25091 &               \\
\hline
\\[-1ex]
\multicolumn{4}{c}{LinSym(5.6)}\\
\\[-1ex]
d2Z &    -0.00125 &   \mcc{$-$} &     37.0 \,\% \\
d3Z &    -0.00493 &    -0.00647 &     34.6 \,\% \\
d4Z &    -0.01008 &    -0.01383 &               \\
d5Z &    -0.01077 &    -0.01151 &               \\
\hline
\end{tabular*}
\end{table}

\begin{table*}
\small
\caption{\ Final estimations of the three-body relativistic and adiabatic corrections compared to the three-body nonrelativistic BO potential.
Numbers in parentheses are uncertainties of the rightmost digits. 
The energy unit is mK.}
\label{tab:post-final}
\begin{tabular*}{\textwidth}{@{\extracolsep{\fill}}ld{2.9}d{2.9}d{2.10}d{2.9}d{3.9}d{6.6}}
\hline\\[-0.7ex]
configuration &
\mcc{$U_3^\mathrm{CG}$} &
\mcc{$U_3^\mathrm{D2}$} &
\mcc{$U_3^\mathrm{oo}$} &
\mcc{$U_3^\mathrm{rel}$} &
\mcc{$U_3^\mathrm{ad}$} &
\mcc{$U_3^\mathrm{BO}$} \\[1ex] 
\hline
Equilat(4) &
  17.8(8) & -7.3(15) & 0.16(9) & 10.7(17) & -76.(23) & -56368.(101)\\
Equilat(5.6) &
  0.040(3) & -0.031(9) & -0.064(5) & -0.056(10) & -0.25(8) & -90.3(5) \\
Equilat(7) &
  -0.00427(16) & 0.00149(21) & -0.01532(9) & -0.0181(3) & 0.015(4) & 16.09(4) \\
Equilat(9) &
  -0.00050(5) & 0.00016(4) & -0.002473(13) & -0.00281(6) & 0.0018(5) & 1.760(4) \\
LinSym(4) &
  -0.646(6) & 0.31(11) & 0.617(15) & 0.28(11) & 4.2(13) & 1818.(12) \\
LinSym(5.6) &
  0.0049(4) & -0.00132(4) & 0.0666(26) & 0.0702(26) & -0.014(4) & -18.58(10) \\
\hline
\end{tabular*}
\end{table*}

To allow for a more thorough comparison of our BO potential with the one from Ref.~\citenum{Cencek:09}, we have randomly chosen 50 triangle configurations considered in this reference and have performed calculations for these configurations using our computation protocol. 
Our results for about 60\% of the considered configurations differ from the results obtained in Ref.~\citenum{Cencek:09} by more than our estimated uncertainty $\,\sigma_3^\mathrm{BO}$.
The largest difference of 6$\,\sigma_3^\mathrm{BO}$ is observed for the configuration with the interatomic distances $(r_{12},r_{23},r_{31})=(4.90, 12.00, 14.30)$~bohr. 
This corresponds to a difference of 1.1\% with respect to the value of the BO potential.
The largest relative difference of 6.7\% with respect to $U_3^\mathrm{BO}$, amounting to 1.8$\,\sigma_3^\mathrm{BO}$, was obtained for the configuration with distances equal to $(3.50, 5.10, 6.90)$~bohr.
Overall, the largest differences occur for configurations with larger interatomic distances as our FCI contribution is more accurate and, as discussed above, the importance of the FCI contribution increases with the size of the system.

For all configurations used in the construction of the three-body BO potential fit, the value of the estimated uncertainty with respect to the value of the potential ranges from less than 0.1\% for small triangles to almost 62\% for the configuration with the interatomic distances of $(5.16, 5.95, 9.73)$~bohr.
Our mean relative uncertainty is 1.2\% and its median is 0.3\%.
Note, that the uncertainty of the three-body BO potential calculated in Ref.~\citenum{Cencek:09} was estimated to be globally at the level of 2\%. 

\subsection{Post-BO corrections}

In the calculations of the post-BO corrections to the three-body potential of helium we adopted a simple one-step procedure. 
All calculations were performed at only one level of theory, using CCSD(T) method in the case of the relativistic corrections and CCSD for the adiabatic correction.
Basis set convergence of the results obtained using our basis sets is presented in Tables~\ref{tab:rel-conv} and \ref{tab:ad-conv}, respectively.
Additionally, we investigated the importance of neglected higher correlation effects by providing relative corrections due to the full iterative treatment of triple excitations.
In the case of the relativistic orbit-orbit correction this data is missing as the Dalton package does not allow for calculations at the CCSDT level of theory.

Basis set convergence of the CCSD(T) results for the relativistic corrections is in most cases smooth and monotonic.
The apparent stabilization of extrapolants occurs early, already for $X\geq4$.
Higher correlation effects are either small, less than 10\% of the CCSD(T) value, or a given correction is negligible compared to the BO potential -- this is the case for the two-electron Darwin correction.
Therefore, as the final values of relativistic corrections we take the d(34)Zu extrapolated results with uncertainty estimated as the absolute difference between the d4Zu value and the d(34)Zu extrapolant.

The extrapolated CCSD results for the adiabatic correction also seem to stabilize for $X\geq4$.
However, in this case the contribution of higher excitations corresponds to approximately 30\% of the CCSD value. 
The configurations with short interatomic distances are exceptions as full iterative triples correspond to only 2.4\% and 5.5\% for Equilat(4) and LinSym(4), respectively.
In the end, as the final values of the adiabatic correction we use the d(34)Z extrapolated results and estimate the uncertainty as 30\% of this value. 

Table~\ref{tab:post-final} collects our recommended values of the post-BO corrections for all test configurations. 
The final value of the relativistic correction $U_3^\mathrm{rel}$ is obtained by summing all the components as defined in Eq.~\eqref{rel:components}, and its uncertainty is calculated as the sum in quadrature of the uncertainties of each component.
Note, that the post-BO corrections for the test configurations are always smaller than the uncertainty of the three-body BO potential.
After performing calculations for the full set of points used to produce final fits of the potentials, we found that the post-BO corrections were larger than the estimated uncertainty of the BO potential for less than 9\% of calculated configurations, mainly for very small triangles.

\section{Fitting of three-body potentials}
\label{sec:fit3b}

\subsection{Form of the fitting functions}

In this work, we provide separate analytic fits for each component of the three-body potential for helium, namely the BO interaction energy and the relativistic and adiabatic corrections. 
Their sum defines the total three-body potential, 
\begin{equation}
U_3=U_3^\mathrm{BO}+U_3^\mathrm{rel}+U_3^\mathrm{ad},
\end{equation}
which is used in Secs.~\ref{sec:virial} and \ref{sec:acoustic} in the determination of thermophysical properties of gaseous helium.
Fitting functions for all the potentials, $U_3^\mathrm{Y}$, $\mathrm{Y}\in\{\mathrm{BO},\;\mathrm{rel},\;\mathrm{ad}\}$, share the same general form, being the sum of a short-range exponentially decaying part, $U_{\mathrm{sr}}^\mathrm{Y}$, and two long-range contributions describing asymptotic behavior of a given potential in two possible fragmentation channels,
\begin{equation}\label{eq:fit-general}
U_3^\mathrm{Y}=U_{\mathrm{sr}}^\mathrm{Y}+U_{\mathrm{3a}}^\mathrm{Y}+U_{\mathrm{a-d}}^\mathrm{Y}.
\end{equation}
The first fragmentation channel, modeled by the $U_{\mathrm{3a}}^\mathrm{Y}$ function, corresponds to the situation when all three atoms move away independently to infinity and all three interatomic distances become arbitrarily large (fragmentation into three isolated atoms).
The second channel, modeled by the $U_{\mathrm{a-d}}^\mathrm{Y}$ function, describes the situation when two atoms stay at a limited, small distance from each other, while the third atom moves to infinity and two interatomic distances become arbitrarily large (fragmentation into an isolated atom and a diatom). 
Only proper inclusion of both channels in the fitting function guarantees a complete theoretical description of the fragmentation processes in the system of three helium atoms.
The analytic form of all three components of Eq.~\eqref{eq:fit-general} is presented in detail in the following sections.
The implementation of the fits for the potentials $U_3^\mathrm{Y}$ in the form of a Fortran 2003 code is included in the ESI.$^\dag$

Geometry dependence of our potentials is expressed in terms of interatomic distances $r_{IJ}$ introduced in Sec.~\ref{sec:calc3b} and cosines of internal angles of the triangle formed by atoms. 
The latter are not independent variables but are related to the interatomic distances through the cosine rule.
For example, $c_1\equiv\cos\theta_1$, where $\theta_1$ is the angle between vectors $\mathbf{r}_3-\mathbf{r}_1$ and $\mathbf{r}_2-\mathbf{r}_1$, can be calculated as
\begin{equation}\label{eq:fit-cosines}
c_1=\frac{r_{31}^2+r_{12}^2-r_{23}^2}{2r_{31}r_{12}}.
\end{equation}
The corresponding formulas for $c_2\equiv\cos\theta_2$ and $c_3\equiv\cos\theta_3$ are obtained by applying the cyclic permutation operators $P_{123}$ and $P_{132}$, respectively, to both sides of Eq.~\eqref{eq:fit-cosines}. 
It is understood that any given operator $P_\pi$ applies permutation $\pi$ to atomic indices 1, 2, and 3 in the expression to the right of $P_\pi$, or, equivalently, permutes the variables $\textbf{r}_1$, $\textbf{r}_2$, and $\textbf{r}_3$.
In the following, by $S_{123}$ we denote the symmetrization operator that includes all possible permutations of three atomic indices,
\begin{equation}
S_{123}=1+P_{12}+P_{23}+P_{13}+P_{123}+P_{132}.
\end{equation}
The symmetrization operator $\tilde{S}_{123}$ is defined by considering only even permutations,
\begin{equation}
\tilde{S}_{123}=1+P_{123}+P_{132}.
\end{equation}

\subsubsection{Short-range part}

The short-range functions $U_{\mathrm{sr}}^\mathrm{Y}$ used by us have the following form 
\begin{equation}\label{eq:fit-SR}
U_{\mathrm{sr}}^\mathrm{Y} = \sum_{l=1}^{L} \; \sum_{\substack{
0 \le i \le j \le k \le N_\mathrm{max} \\ 
i+j+k \le N_\mathrm{sum}
}} A_{l,ijk} \: S_{123} \: r_{12}^{i} r_{23}^{j} r_{31}^{k} \,
e^{-\alpha_l r_{12}-\beta_l r_{23}-\gamma_l r_{31}},
\end{equation} 
where $A_{l,ijk}$ are linear parameters and $\alpha_l$, $\beta_l$, and $\gamma_l$ are nonlinear parameters to be fitted. 
As the summation limit $L$, we chose $L=4$ for the three-body BO potential and $L=3$ for the post-BO corrections. 
In all three cases $N_\mathrm{max}=3$ and $N_\mathrm{sum}=7$ were used.
Consequently, there are 72 (54) linear and 12 (9) nonlinear fitting parameters in the $U_{\mathrm{sr}}^\mathrm{BO}$ ($U_{\mathrm{sr}}^\mathrm{rel/ad}$) function.

The form of the short-range function in Eq.~\eqref{eq:fit-SR} differs from the exponentially decaying terms employed in Refs.~\citenum{cencek2007three,Cencek:09,cencek2013three} in the studies on three-body potentials for helium and argon.
In these works, the perimeter of the triangle formed by atoms was used as the main argument of the exponential functions.
The resulting fitting function is then unable to correctly distinguish linear geometries when the positions of the two outer atoms are fixed and the middle atom moves between them, because the sum of all interatomic distances remains constant. 
By contrast, all the exponential terms used in Eq.~\eqref{eq:fit-SR} change during this movement.
Another difference is that we describe the angular dependence in terms of powers of interatomic distances $r_{IJ}$, rather than in terms of Legendre polynomials of cosines of internal angles $c_I$ as used in Refs.~\citenum{cencek2007three,Cencek:09,cencek2013three}. 
After extensive testing, we found that the direct use of interatomic distances provides superior fits in comparison with previous approaches. 

\subsubsection{Three-atomic fragmentation channel}

The asymptotic expansion of the three-body potentials in the three-atomic fragmentation channel can be rigorously derived by applying the symmetry-adapted perturbation theory and neglecting exchange and charge-overlap effects. \cite{Lotrich:97}
This results in a triple series in the inverse powers of interatomic distances, $r_{12}^{-l}$, $r_{23}^{-m}$, and $r_{31}^{-n}$, the leading terms of which have been known for a long time. \cite{Axilrod:43,Muto:43,Bell:70,Doran:71} 
Terms with a given value of the sum of the exponents, $s=l+m+n$, will be for brevity referred to as terms vanishing as $R^{-s}$. 
The coefficient at each product $r_{12}^{-l}r_{23}^{-m}r_{31}^{-n}$ can be represented by the product of the ``dynamic'' constants $Z$, depending only on the electric properties of interacting atoms, and the geometry dependent angular factor $W\equiv W(c_1,c_2,c_3)$. 
To eliminate the singular behavior of the functions $U_\mathrm{3a}^\mathrm{Y}$ when one of the interatomic distances approaches zero, we replaced every occurrence of the factor $r_{IJ}^{-k}$ by the function $d_k(\eta,r_{IJ})$ defined as
\begin{equation}\label{eq:fit-d}
d_k(\eta,r) = \frac{f_{k+1}(\eta r)}{r^k},
\end{equation}
where $\eta$ is a damping parameter to be fitted, and $f_n(x)$ is the Tang-Toennies damping function, \cite{tang1984improved}
\begin{equation}\label{TTdamp}
f_n(x) = 1 - e^{-x} (1 + x + x^2/2! + \dots + x^n/n!).
\end{equation}
Note, that for small $r$ the function $d_k(\eta,r)$ vanishes at $r=0$ as $\sim\! r^2$. 
To simplify the following presentation, we introduce another function defined in terms of the $d_k(\eta,r)$ factors,
\begin{equation}
D(\eta;l,m,n)=d_l(\eta,r_{12})\,d_m(\eta,r_{23})\,d_n(\eta,r_{31}).
\end{equation}
If any of the $l$, $m$, or $n$ arguments is zero then the corresponding $d_k(\eta,r)$ factor is replaced by unity.

In the function $U_\mathrm{3a}^\mathrm{BO}$ we included all terms vanishing as $R^{-15}$ that originate in the third and fourth order of the perturbation theory,
\begin{equation}
U_\mathrm{3a}^{\mathrm{BO}} = U^{(3)}_\mathrm{3a} + U^{(4)}_\mathrm{3a}.
\end{equation}
As the leading fifth-order terms vanish as $R^{-15}$, our expansion provides a complete representation of the three-body BO potential in the three-atomic fragmentation channel for all terms vanishing as $R^{-14}$ or slower.

The third-order contribution is \cite{Bell:70,tang2012long,MPunp}
\begin{equation}\label{eq:fit-BO3}
\begin{split}
U^{(3)}_\mathrm{3a}
&=                    Z_{111}\,W_{111} \: D(\eta^{(3)};3,3,3)
 + \tilde{S}_{123} \: Z_{112}\,W_{112} \: D(\eta^{(3)};3,4,4) \\
&+ \tilde{S}_{123} \: Z_{122}\,W_{122} \: D(\eta^{(3)};4,5,4) 
 + \tilde{S}_{123} \: Z_{113}\,W_{113} \: D(\eta^{(3)};3,5,5) \\
&+                    Z_{222}\,W_{222} \: D(\eta^{(3)};5,5,5)
 + \tilde{S}_{123} \: Z_{114}\,W_{114} \: D(\eta^{(3)};3,6,6) \\
&+         S_{123} \: Z_{123}\,W_{123} \: D(\eta^{(3)};4,6,5). 
\end{split}
\end{equation}
The values of the asymptotic constants $Z_{ijk}$ and the form of the angular factors $W_{ijk}$ are presented in Appendix~\ref{app:Z} and \ref{app:W}, respectively. 
To the best of our knowledge, the last two terms in Eq.~\eqref{eq:fit-BO3} have not been considered previously in the literature.

The fourth-order contribution is
\begin{equation}\label{eq:fit-BO4}
\begin{split}
U^{(4)}_\mathrm{3a} 
&=         S_{123} \: Z_1\,W_1 \: D(\eta^{(4)};7,3,4) 
 + \tilde{S}_{123} \: Z_2\,W_2 \: D(\eta^{(4)};6,4,4) \\
&+ \tilde{S}_{123} \: Z_3\,W_3 \: D(\eta^{(4)};8,3,3) 
 + \tilde{S}_{123} \: Z_4\,W_4 \: D(\eta^{(4)};8,3,3) \\
&+ \tilde{S}_{123} \big(Z_5\,P_0(c_2)+Z_6\,P_2(c_2)\big) \, D(\eta^{(4)};6,6,0) \\
&+         S_{123} \big(Z_7\,P_0(c_2)+Z_8\,P_2(c_2)\big) \, D(\eta^{(4)};8,6,0) \\
&+ \tilde{S}_{123} \big(Z_9\,P_1(c_2)+Z_{10}\,P_3(c_2)\big) \, D(\eta^{(4)};7,7,0),
\end{split}
\end{equation}
where the coefficients $Z_i$, $i\in \{1,\cdots,10\}$, are treated as parameters to be fitted, the angular factors $W_i$, $i\in \{1,\cdots,4\}$, are presented in Appendix~\ref{app:W}, and $P_l(x)$ are the Legendre polynomials of order $l$. 
The fourth-order contribution to the long-range expansion of the three-body BO potential has been previously published in Ref.~\citenum{Lotrich:97} with minor corrections made in Ref.~\citenum{cencek2007three}.
The formulas for $U^{(4)}_\mathrm{3a}$ presented in this reference differ from ours and we believe that they are partially incorrect. \cite{MPunp}

Asymptotic expansions of the post-BO potentials can be split into two contributions of different origins, \cite{Cencek:12,MPunp}
\begin{equation}
U^{\mathrm{Y}}_\mathrm{3a}=U^{\mathrm{Y},A}_\mathrm{3a}+U^{\mathrm{Y},E}_\mathrm{3a},
\end{equation}
where $\mathrm{Y}\in\{\mathrm{rel},\;\mathrm{ad}\}$. 
The first contribution, $U^{\mathrm{Y},A}_\mathrm{3a}$, collects all terms that result solely from the first-order corrections to the wave functions of each interacting atom due to the intra-atomic part of the operator defining a given post-BO correction Y. 
The second contribution, $U^{\mathrm{Y},E}_\mathrm{3a}$, collects the remaining terms whose origin is specific to the correction Y. 

In our $U^{\mathrm{Y},A}_\mathrm{3a}$ functions we include only the leading terms that vanish with the interatomic distances as $R^{-9}$,
\begin{equation}\label{eq:fit-YA}
U^{\mathrm{Y},A}_\mathrm{3a}= Z_{111}^{\mathrm{Y},A}\,W_{111} \: D(\eta^\mathrm{Y};3,3,3),
\end{equation}
where the values of the constants $Z_{111}^{\mathrm{Y},A}$ are presented in Appendix~\ref{app:Z}.

In the case of the relativistic correction, the additional terms come from nontrivial multipole expansion of the inter-atomic part of the orbit-orbit operator $H^\mathrm{oo}$ in Eq.~\eqref{rel:oo}. \cite{Meath:66,Meath:68,MPunp}
After including all terms vanishing as $R^{-9}$ or slower, our $U^{\mathrm{rel},E}_\mathrm{3a}$ function has the form
\begin{equation}\label{eq:fit-YE}
\begin{split}
U^{\mathrm{rel},E}_\mathrm{3a} 
&= \tilde{S}_{123} \: Z_1^{\mathrm{rel},E} \, W_1^{\mathrm{rel},E} \: 
D(\eta^\mathrm{rel};1,3,3) \\
&+ \tilde{S}_{123} \: Z_{2,3}^{\mathrm{rel},E} \, W_2^{\mathrm{rel},E} \: 
D(\eta^\mathrm{rel};1,4,4) \\
&+         S_{123} \: Z_{2,3}^{\mathrm{rel},E} \, W_3^{\mathrm{rel},E} \: 
D(\eta^\mathrm{rel};3,2,4) \\
&- \frac35 \: Z_4^{\mathrm{rel},E} \, W_{111} \: 
D(\eta^\mathrm{rel};3,3,3),
\end{split}
\end{equation}
where we use the same fitted damping parameter $\eta^\mathrm{rel}$ in both Eq.~\eqref{eq:fit-YA} and Eq.~\eqref{eq:fit-YE}.
The values of the asymptotic constants $Z_1^{\mathrm{rel},E}$, $Z_{2,3}^{\mathrm{rel},E}$, and $Z_4^{\mathrm{rel},E}$, together with the form of the angular factors $W_1^{\mathrm{rel},E}$, $W_2^{\mathrm{rel},E}$, and $W_3^{\mathrm{rel},E}$ are presented in Appendix~\ref{app:Z} and \ref{app:W}, respectively.

In the case of the adiabatic correction, the additional terms appear due to the presence of derivatives with respect to nuclear coordinates in the $H^\mathrm{ad}$ operator, Eq.~\eqref{ad:tot}. \cite{Przybytek12ad,MPunp}
The leading terms of this type vanish as $R^{-11}$.
Therefore, we may set
\begin{equation}
U^{\mathrm{ad},E}_\mathrm{3a}=0
\end{equation}
to assure a complete description of the asymptotic expansion of the adiabatic correction up to terms vanishing as $R^{-9}$. 

To summarize, our $U_\mathrm{3a}^\mathrm{BO}$ function has 10 linear parameters and 2 nonlinear damping parameters to be fitted. 
Each of the functions modelling the behavior of the post-BO corrections has only one fitted damping parameter.

\subsubsection{Atom-diatom fragmentation channel}

To model the asymptotic behavior of the three-body potentials in the atom-diatom fragmentation channel, it is convenient to describe the geometry of a system through the set of Jacobi coordinates.
Let us assume, for example, that the diatom is formed by atoms 1 and 2, while atom 3 is treated as the one that separates from the diatom. 
The Jacobi coordinates are then: the length of the vector $\mathbf{r}_{12}=\mathbf{r}_2-\mathbf{r}_1$ connecting atoms in the diatom, the length of the vector $\mathbf{R}_{12}=\mathbf{r}_3-(\mathbf{r}_1+\mathbf{r}_2)/2$ connecting the barycenter of the diatom with the separating atom, and the angle $\theta_{12}$ between the vectors $\mathbf{r}_{12}$ and $\mathbf{R}_{12}$.
In terms of the interatomic distances, the Jacobi coordinates can be expressed as $r_{12}$, $R_{12}=1/2\sqrt{2(r_{31}^2+r_{23}^2)-r_{12}^2}$, and $c_{12}\equiv\cos{\theta_{12}}=(r_{31}^2-r_{23}^2)/(2r_{12}R_{12})$, respectively.
The long-range behavior of the atom-diatom potentials can be then represented as a series in inverse powers of $R_{12}$, where the expansion coefficients include Legendre polynomials in $c_{12}$. \cite{Buckinghan:67,Pack:76,Cvitas2006}
Again, we remove the singularity at $R_{12}=0$ by using the damping function defined in Eq.~\eqref{eq:fit-d}.

In a system of three identical atoms, all three separation scenarios must be treated equally.
Therefore, to model the second fragmentation channel, we employ a function of the form
\begin{equation}\label{eq:fit-ad}
U_{\mathrm{a-d}}^\mathrm{Y} = \tilde{S}_{123} \sum_{m=1}^M
 \sum_{i=0}^2 a_{m,i}\, r_{12}^i e^{-\zeta_r r_{12}} \,
d_{n_m}(\zeta_R,R_{12}) \,
P_{l_m}(c_{12}),
\end{equation}
where $a_{m,i}$ are linear parameters and $\zeta_r$ and $\zeta_R$ are nonlinear damping parameters to be fitted. 
Fixed values of $n_m$ and $l_m$ correspond to the powers of $R_{12}^{-1}$ and the orders of the Legendre polynomials, respectively. 
In the case of the BO potential, we include $M=5$ pairs of $(n_m,l_m)$ values, $(n_m,l_m) \in \{(6,0),(6,2),(8,0),(8,2),(8,4)\}$. 
In the fit for the relativistic correction we use $M=4$ with $(n_m,l_m)\in\{(4,0),(4,2),(6,0),(6,2)\}$, and our fit for the adiabatic correction has $M=2$ pairs of parameters $(n_m,l_m)\in\{(6,0),(6,2)\}$.
The inclusion of the parameter pairs with $n_m=4$ in $U_{\mathrm{a-d}}^\mathrm{rel}$ is again necessary to account for the terms that appear in the long-range asymptotics due to the multipole expansion of the interatomic part of the orbit-orbit operator.
Note, that the presence of the $e^{-\zeta_r r_{12}}$ factor in Eq.~\eqref{eq:fit-ad} is crucial to ensure that the function modelling the atom-diatom fragmentation channel does not interfere with the asymptotic expansions of the potentials in the three-atomic channel. \cite{Lang:23alpha}

In the end, our $U_{\mathrm{a-d}}^\mathrm{BO}$, $U_{\mathrm{a-d}}^\mathrm{rel}$, and $U_{\mathrm{a-d}}^\mathrm{ad}$ functions have 15, 12, and 6 linear parameters to be fitted, respectively.
The number of fitted damping parameters is two in each case.

\subsection{Generation of the set of fitting points}

Effective selection of grid points is an important problem in fitting potential energy surfaces. \cite{Metz:2020}
In our work the set of grid points was generated using the iterative approach from Ref.~\citenum{Lang:23alpha}.
As for a given atomic configuration the value of the three-body BO potential is 2 to 4 orders of magnitude larger than the post-BO corrections, cf.\ Table~\ref{tab:post-final}, the procedure was performed using only BO energies. 

The initial set of 286 points was created by combining 50 points randomly sampled from the dataset of Ref.~\citenum{Cencek:09}, 103 points handpicked to investigate configurations of interest (mostly linear and equilateral ones), and a set of 130 points satisfying the conditions $\min(r_{12},r_{23},r_{31}) > 1.2$~bohr and $\max(r_{12},r_{23},r_{31}) < 14$~bohr created through the Sobol sequence.
In each step of the iterative procedure, $\sim\!\!10^6$ sets of 16 nonlinear parameters appearing in our functional form of the BO potential were randomly generated in a suitably chosen 16D box.
Subsequently, the current set of grid points was used in a local optimization of the nonlinear parameters to obtain $\sim\!\!10^6$ independent fitting functions. 
The optimization was performed using the trust-region dogleg procedure \cite{Powell1970} as implemented in the GSL library. \cite{gsl_lib}
During the optimization, the linear parameters (for fixed values of the nonlinear ones) were obtained using the weighted linear least-squares minimization with weights equal to the inverse square of the calculated uncertainties, $1/(\sigma_3^\mathrm{BO})^2$. 
The fits prepared this way were sorted according to their mean and maximum absolute relative errors, and the top 20 fits were picked for further analysis. 
The analysis consisted of evaluating the selected 20 fits for $\sim\!\!10^4$ randomly distributed trimer configurations and finding regions of the configuration space where the largest differences between the fit values were observed. 
A new batch of 40--80 grid points covering the problematic regions were constructed, the corresponding values of the three-body BO potential were computed as described in Sec.~\ref{sec:calc3b}, and the whole procedure was repeated with the enlarged set of grid points. 
The generation process was finished after 3 iterations when the differences between trial fits were acceptably small.

The final set of grid points consisted of 463 configurations. 
This set was then split into two sets, training and testing, comprising 414 and 49 points, respectively.
Only the training set was used to produce final fits for the potentials $U_3^\mathrm{Y}$, $\mathrm{Y}\in\{\mathrm{BO},\;\mathrm{rel},\;\mathrm{ad}\}$, while both sets were used to assess the correctness of the fits.
Each fit was constructed using a procedure similar to a single step of the iterative process described above, that is, out of $\sim\!10^6$ functions obtained by optimizing randomly generated initial values of nonlinear parameters, the one with the smallest mean and maximum absolute relative error was selected.
During optimization of linear parameters, appropriate theoretical uncertainties $\sigma_3^\mathrm{Y}$ were employed to construct the weights in the weighted linear least-squares minimization.

\subsection{Fit of BO potential}

\begin{figure}
\centering
\includegraphics[width=1\columnwidth]{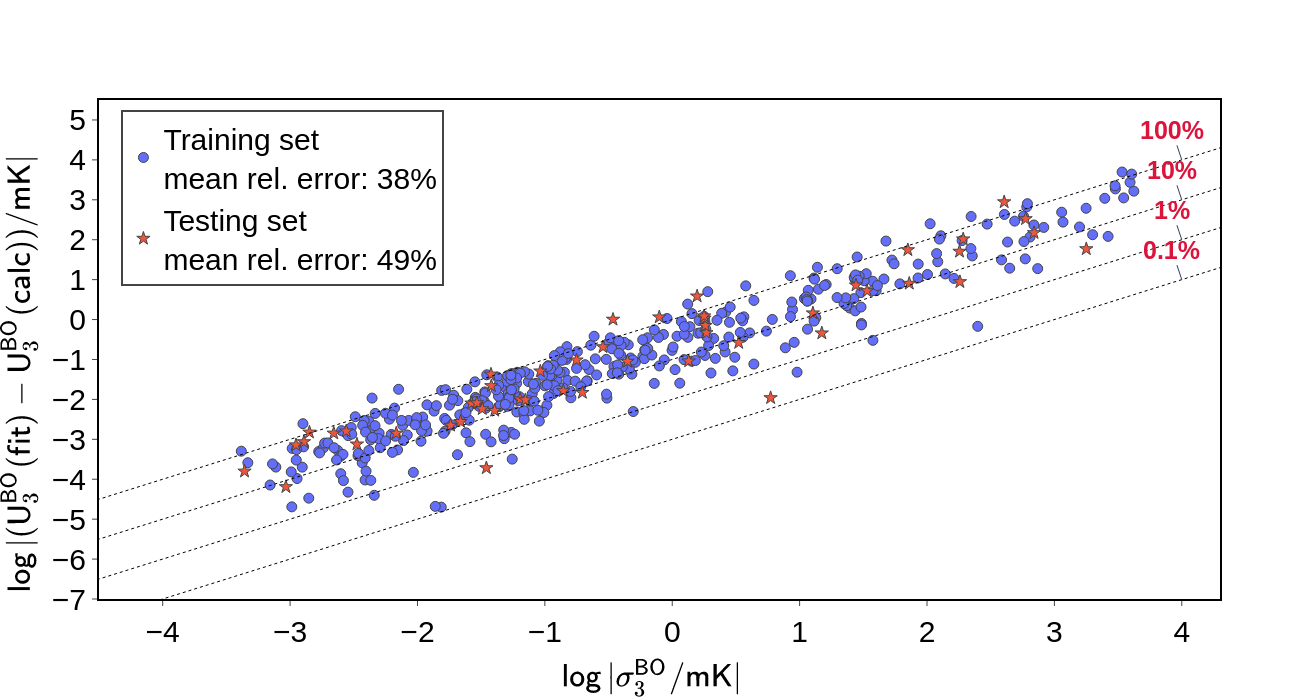}
\caption{
Absolute differences between the three-body BO potential predicted from the fit, $U_3^\mathrm{BO}(\mathrm{fit})$, and obtained from \emph{ab initio} calculations, $U_3^\mathrm{BO}(\mathrm{calc})$, versus estimated theoretical uncertainties $\sigma_3^\mathrm{BO}$.
The dotted lines and percentage in the legend correspond to relative errors with respect to the estimated uncertainty (1$\sigma_3^\mathrm{BO}=100\%$).}
\label{fig:fitBO}
\end{figure}

Our final fit of the three-body BO potential $U_3^\mathrm{BO}$ has the mean absolute relative error with respect to the theoretical uncertainties of 0.39$\,\sigma_3^\mathrm{BO}$ and its median is 0.26$\,\sigma_3^\mathrm{BO}$. 
This corresponds to the mean absolute percentage error with respect to the \emph{ab initio} data of 0.52\% with median 0.08\%.
The largest deviation of 2.63$\,\sigma_3^\mathrm{BO}$ is observed for the configuration with the interatomic distances $(r_{12},r_{23},r_{31})=(1.6,8.4,10.0)$~bohr, where the sum of the BO pair energies is five orders of magnitude larger than the non-additive three-body contribution.
Two other configurations with the error larger than 2$\,\sigma_3^\mathrm{BO}$ are triangles with the shortest side length smaller than 4~bohr. 
In Fig.~\ref{fig:fitBO}, we present the differences between the calculated \emph{ab initio} and fit-predicted values of the three-body BO potential compared to the values of the estimated uncertainties. 

\begin{figure}[t]
\centering
\includegraphics[width=1\columnwidth]{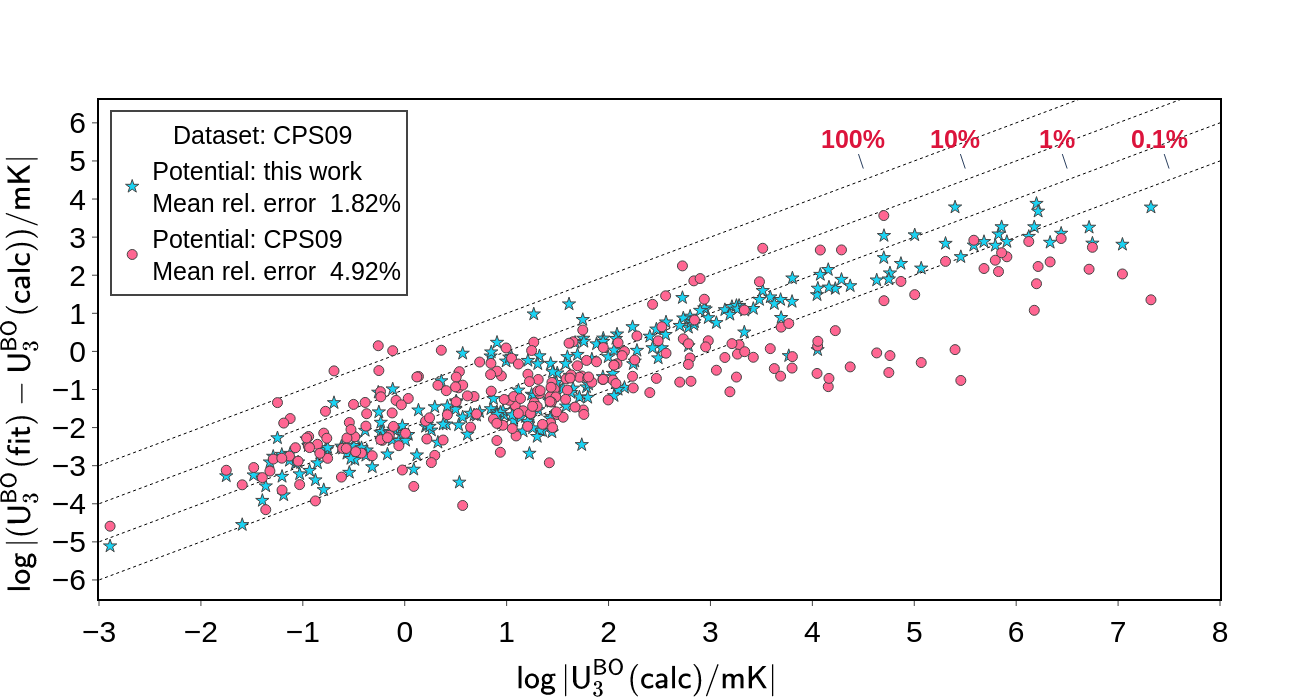}
\caption{
Comparison of the CPS09 potential from Ref.~\citenum{Cencek:09} and the $U_3^\mathrm{BO}$ fit obtained in this work on the original CPS09 dataset.
The percentage in the legend corresponds to relative errors with respect to the calculated energy.}
\label{fig:comp2009}
\end{figure}

\begin{figure}
\centering
\includegraphics[width=1\columnwidth]{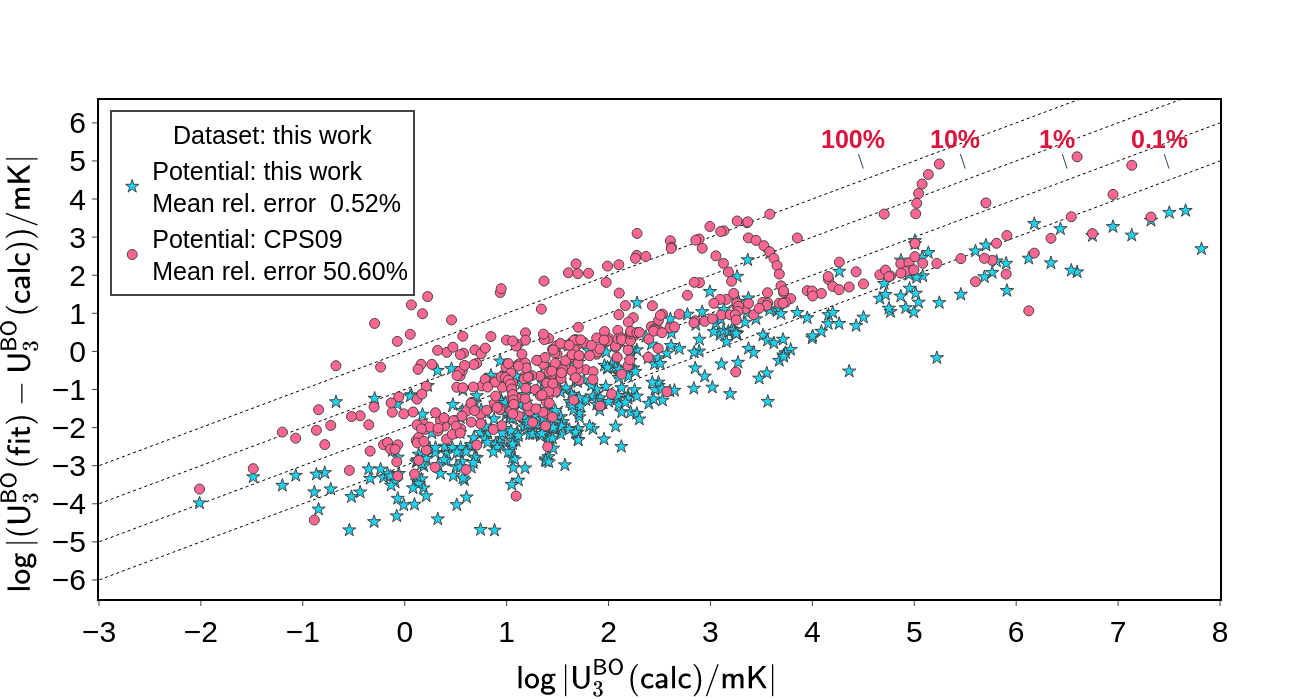}
\caption{
Comparison of the CPS09 potential from Ref.~\citenum{Cencek:09} and the $U_3^\mathrm{BO}$ fit obtained in this work on the dataset from this work.
The percentage in the legend corresponds to relative errors with respect to the calculated energy.}
\label{fig:comp2022}
\end{figure}

To show an improvement in the description of the three-body BO potential of helium achieved in this work, we compare our fit with the previous fit of Cencek \emph{et al.} \cite{Cencek:09} denoted as the CPS09 potential. 
In Fig.~\ref{fig:comp2009}, we compare the performance of both fits on the CPS09 dataset calculated in Ref.~\citenum{Cencek:09}.
Our fit provides similar results to the CPS09 potential even though, besides 50 shared configurations, it was not fitted on the same dataset.
Overall, our present fit provides two times smaller mean fit error than the CPS09 fit. 
In Fig.~\ref{fig:comp2022}, we compare the performance of both fits on the dataset from this work. 
Here, the performance of our new fit is better by two orders of magnitude in terms of mean fit error. 
This improvement comes mainly from better description of linear configurations.
Five chains of red circles that stand out in Fig.~\ref{fig:comp2022} correspond to linear configurations with the distance between two outer atoms fixed at 6, 7.5, 8, 10, and 12~bohrs. 
While the CPS09 fit predicts fairly well the situation where the third atom is precisely in the middle (the lowest point in the chain), it fails when the third atom approaches either of the outer ones. 
Moreover, the CPS09 potential does not provide a proper description of the atom-diatom fragmentation channel. 
Configurations for which values predicted from the CPS09 potential have error exceeding $100\%$ are the ones in which one of the atoms is at large distance from the other two.

\subsection{Fits of post-BO corrections}

\begin{figure}[t]
\centering
\includegraphics[width=1\columnwidth]{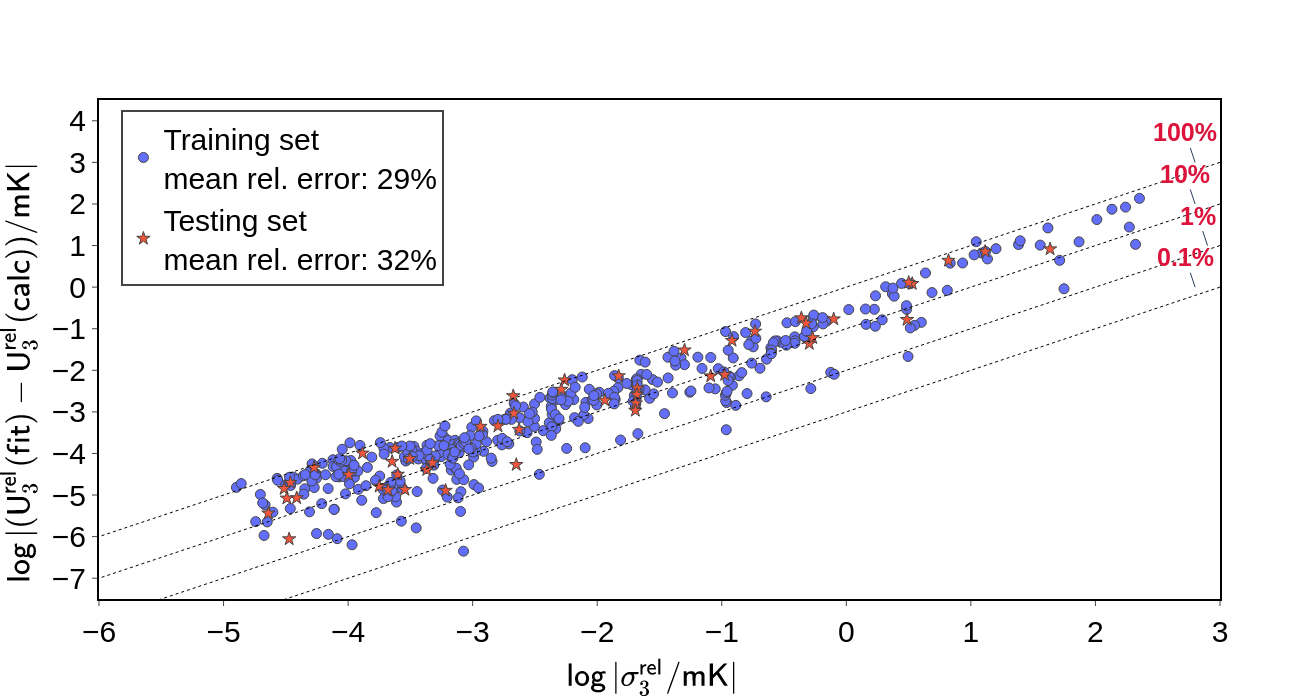}
\caption{
Absolute differences between the three-body relativistic correction predicted from the fit, $U_3^\mathrm{rel}(\mathrm{fit})$, and obtained from \emph{ab initio} calculations, $U_3^\mathrm{rel}(\mathrm{calc})$, versus estimated theoretical uncertainties $\sigma_3^\mathrm{rel}$.
The dotted lines and percentage in the legend correspond to relative errors with respect to the estimated uncertainty (1$\sigma_3^\mathrm{rel}=100\%$).}
\label{fig:fitRel}
\end{figure}

\begin{figure}[t]
\centering
\includegraphics[width=1\columnwidth]{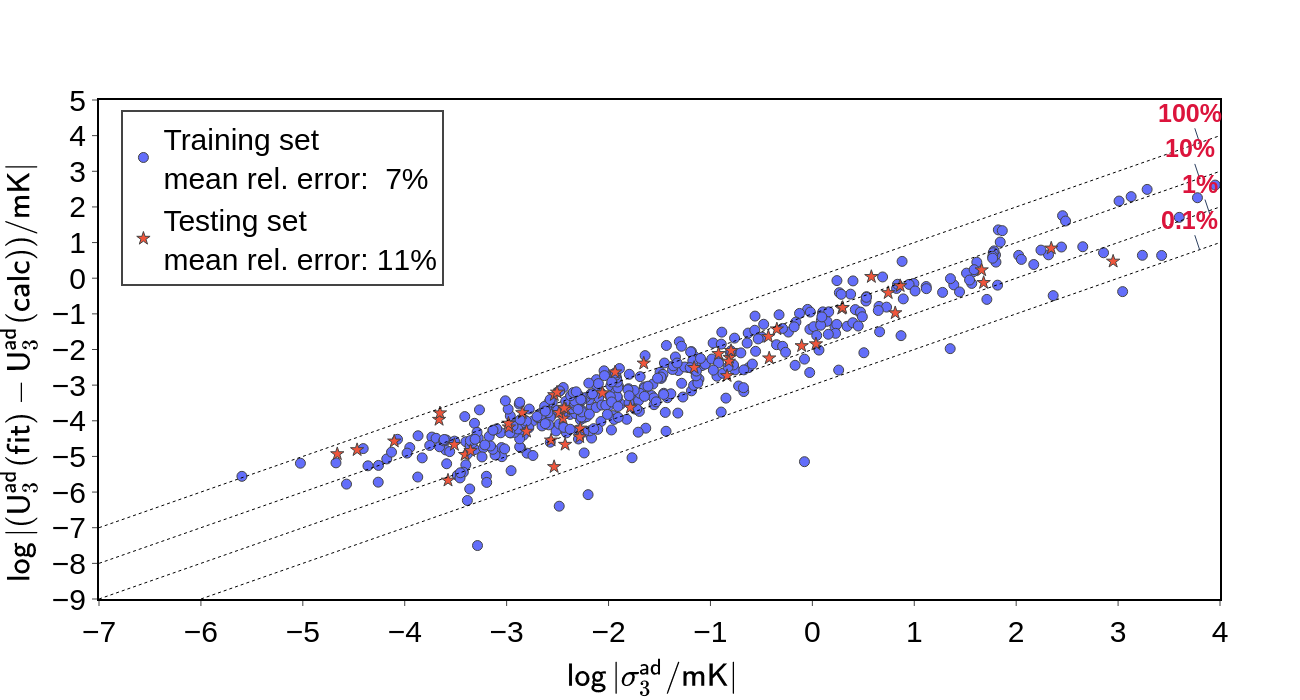}
\caption{
Absolute differences between the three-body adiabatic correction predicted from the fit, $U_3^\mathrm{ad}(\mathrm{fit})$, and obtained from \emph{ab initio} calculations, $U_3^\mathrm{ad}(\mathrm{calc})$, versus estimated theoretical uncertainties $\sigma_3^\mathrm{ad}$.
The dotted lines and percentage in the legend correspond to relative errors with respect to the estimated uncertainty (1$\sigma_3^\mathrm{ad}=100\%$).}
\label{fig:fitAD}
\end{figure}

In Figs.~\ref{fig:fitRel} and \ref{fig:fitAD}, we present the differences between the calculated \emph{ab initio} and fit-predicted values of the three-body relativistic and adiabatic correction, respectively, compared to the values of the estimated uncertainties. 

Our final fit of the relativistic correction $U_3^\mathrm{rel}$ has the mean absolute relative error of 0.29$\,\sigma_3^\mathrm{rel}$ and its median is 0.22$\,\sigma_3^\mathrm{rel}$. 
This corresponds to the mean absolute percentage error of 3.78\% with median 0.93\%.
The largest error is 1.75$\,\sigma_3^\mathrm{rel}$ and, similarly as in the case of the BO potential, is observed for a configuration with the shortest interatomic distance smaller than 4~bohr. 

Final fit of the adiabatic correction $U_3^\mathrm{ad}$ has the mean absolute relative error of 0.08$\,\sigma_3^\mathrm{ad}$ and its median is 0.05$\,\sigma_3^\mathrm{ad}$, which translates to the mean absolute percentage error of 2.33\% with median 1.37\%. 
In this case most significant discrepancies are observed mainly for large triangles. 
For instance, the largest error of 1.09$\,\sigma_3^\mathrm{ad}$ is for the equilateral triangle with the side length equal to 16~bohr. 
However, such behavior is expected as the \emph{ab initio} values of the adiabatic correction were calculated at low level of theory, while our fitting function provides terms properly describing asymptotic expansion of this correction in the three-atomic fragmentation channel.
This long-range expansion is then a dominating contribution to the values of the adiabatic correction for large triangles.
Moreover, the fact that errors of the fit do not exceed significantly $\sigma_3^\mathrm{ad}$ for such configurations suggests that our estimation of the magnitude of higher correlation effects was realistic.

\subsection{Fit of local uncertainties}

In order to estimate the uncertainties of physical properties of the helium gas obtained using our three-body potential $U_3$, we provide a fit for a function $\sigma_3$ representing estimated total uncertainties of $U_3$ due to the uncertainty of the \emph{ab initio} calculations. 
The exact value of the total three-body potential is expected to be contained in the range $U_3 \pm \sigma_3$. 
Note that the function $\sigma_3$ is not intended to precisely fit our estimated theoretical uncertainties but rather should fulfill two conditions.
Firstly, it should follow the trends in the behavior of the uncertainties with respect to the system's geometry.
Secondly, it should provide an upper bound to the theoretical errors. 
For the majority ($92\%$) of the configurations in our dataset, the combined contribution of the post-BO corrections is much smaller than the uncertainty of the BO potential.
Therefore, it is justified to assume that the total uncertainty is dominated by the uncertainty of the latter, and only the estimated values of $\sigma_3^\mathrm{BO}$ were used in the fitting.

The functional form of $\sigma_3$ used by us is as follows
\begin{equation}
\sigma_3 = \sum_{l=1}^{2}
S_{123} \: e^{-\alpha_l r_{12}-\beta_l r_{23}-\gamma_l r_{31}}
\sum_{\substack{
0 \le i \le j \le k \le 2 \\ 
i+j+k \le 4
}} A_{l,ijk} \: S_{123} \: r_{12}^{i} r_{23}^{j} r_{31}^{k},
\end{equation} 
where 
$A_{l,ijk}$ are linear parameters (16 in total) and $\alpha_l$, $\beta_l$, and $\gamma_l$ are nonlinear parameters (6 in total) to be fitted. 
Due to the large variance of $\sigma_3^\mathrm{BO}$ across our dataset, we selected a subset of points that was actually used in the fitting. 
We discarded configurations with the estimated uncertainties significantly smaller than the uncertainties for the neighboring ones. 
We further discarded configurations with $\min(r_{12},r_{23},r_{31}) > 14$~bohr as the proper inclusion of three-atomic asymptotic expansions of the potentials in our fits is expected to provide correct results in this case. 
Finally, we discarded configurations with $\min(r_{12},r_{23},r_{31}) < 2$~bohr which are of little importance in the calculations of physical properties for temperatures below 5000~K.
The final subset contained about $85\%$ of all configurations.
The fitting of the $\sigma_3$ function on the selected subset of points was performed using the least-squares procedure. 

The average ratio of the values predicted from the $\sigma_3$ function to the actual estimated uncertainties is $1.8$ with median equal to $1.7$. 
With respect to the values of the three-body BO potential, the $\sigma_3$ function is on average 2.2\% with median 0.45\%. 
While the average is slightly larger than the assumed uncertainty of the CPS09 dataset, equal to 2\%,\cite{Cencek:09} our uncertainty is highly local. 
The largest uncertainties appear only for configurations where all three interatomic distances are small and the median of the fit for $\sigma_3$ is much smaller than 2\%. 

\section{Third pressure virial coefficients}
\label{sec:virial}

The general expression for the third pressure virial coefficient has the form \cite{hirschfelder1954molecular}
\begin{equation}\label{eq:c_virial}
C(T) = 
-\frac{Z_3 - 3 Z_2 Z_1 + 2 Z_1^3}{3V}
+\frac{(Z_2 - Z_1^2)^2}{V^2},
\end{equation}
where 
\begin{equation}
Z_N(V,T) = N!\frac{Q_N(V,T)\;V^N}{Q_1(V,T)^N},
\end{equation}
$V$ is the volume, and $Q_N(V,T)$ is the canonical partition function of a system of $N$ particles at temperature $T$.
This expression is valid for both classical and quantum mechanical calculations and both approaches differ in the definition of $Q_N(V,T)$ and $Z_N(V,T)$.

\subsection{Classical and semiclassical calculations}

In the case of the classical approach, $Z_N(V,T)$ has the form
\begin{equation}
Z_N^\mathrm{class}(V,T) = \int e^{-\beta V_N} ~ d\mathbf{r}_1\cdots d\mathbf{r}_N,
\end{equation}
where $\beta = 1/k_BT$ and $V_N$ is the total potential energy of a configuration with $N$ particles at positions $\mathbf{r}_1,\cdots,\mathbf{r}_N$, defined as the difference between the energy of the $N$-particle system and the sum of energies of $N$ separate particles.
After substitution of this classical partition function into the definition of the third virial coefficient in Eq.~\eqref{eq:c_virial} and integrating over center-of-mass coordinates and Euler angles, we obtain the classical expression for $C(T)$ which can be represented as the sum of two contributions -- the first depending only on the pair potential $U_2(r_{IJ})$ and the second depending also on the three-body potential $U_3(r_{12},r_{23},r_{31})$, \cite{hirschfelder1954molecular}
\begin{equation}\label{eq:C_class}
C^\mathrm{class}(T) = C_\mathrm{add}^\mathrm{class}(T) + C_\mathrm{non-add}^\mathrm{class}(T),
\end{equation}
where
\begin{equation}\label{eq:C_add}
\begin{split}
C_\mathrm{add}^\mathrm{class}(T) &
=-\frac{8\pi^2}{3}\int dr_{12}\,dr_{23}\,d\cos{\theta_2}\,r_{12}^2\,r_{23}^2 \\
&\times
\left(e^{-\beta\,U_2(r_{12})}-1\right)
\left(e^{-\beta\,U_2(r_{23})}-1\right)
\left(e^{-\beta\,U_2(r_{31})}-1\right),
\end{split}
\end{equation}
and
\begin{equation}\label{eq:C_nonadd}
\begin{split}
C_\mathrm{non-add}^\mathrm{class}(T) &
=-\frac{8\pi^2}{3}\int dr_{12}\,dr_{23}\,d\cos{\theta_2}\,r_{12}^2\,r_{23}^2 \\
&\times
\left(e^{-\beta\,V_3(r_{12},r_{23},r_{31})}-1\right)
\,e^{-\beta \left(U_2(r_{12})+U_2(r_{23})+U_2(r_{31})\right)}.
\end{split}
\end{equation}
$V_3(r_{12},r_{23},r_{31})$ in the above equation, defined in terms of the pair and the three-body potential, is $V_3(r_{12},r_{23},r_{31})=U_3(r_{12},r_{23},r_{31})+U_2(r_{12})+U_2(r_{23})+U_2(r_{31})$. 

The classical approach is accurate only for heavy atoms or for very high temperatures. 
In the case of helium gas the classical approach fails for temperatures as high as 500~K. 
Currently, two possible semiclassical methods have been published in the literature to correct the classical approach, the semiclassical expansion of Kihara \cite{kihara1953virial,kihara1955virial} and Yokota, \cite{yokota1960expansion} and the quadratic Feynman--Hibbs (QFH) effective potential approach. \cite{feynman1965pathintegrals}
Kihara and Yokota derived the asymptotic expansion of $C(T)$ in powers of the Planck constant, while the QFH approach only modifies the pair potential to the form
\begin{equation}\label{qfh_pot}
U_2^\mathrm{QFH}(r_{IJ}) 
= U_2(r_{IJ}) 
+ \frac{\hbar^2\beta}{12m}\left(
 \frac{\partial^2U_2(r_{IJ})}{\partial r_{IJ}^2}
+\frac{2}{r_{IJ}}\frac{\partial U_2(r_{IJ})}{\partial r_{IJ}}
\right),
\end{equation}
and $U_2^\mathrm{QFH}(r_{IJ})$ is used instead of $U_2(r_{IJ})$ in Eqs.~\eqref{eq:C_add} and \eqref{eq:C_nonadd}. 
This makes the QFH approach simpler to implement but theoretically less accurate. 
However, as the contribution depending only on the pair potential is dominant -- for instance $C_\mathrm{add}^\mathrm{class}(T)$ for helium represents more than 95\% of $C^\mathrm{class}(T)$ -- Shaul \emph{et al.} \cite{shaul2012path,shaul2012semiclassical} used the QFH approach to obtain very accurate results on par with full quantum calculations down to 50~K. 
Therefore, we use QFH in this work to correct the classical results as well. 

\subsubsection{Computational details}

In the calculations of the two-body contributions we used the pair potential from Ref.~\citenum{czachorowski2020second}.
The uncertainties in the classical third virial coefficient calculations were estimated through the propagation of errors for both the two-body and three-body interaction potentials. 
We evaluated Eq.~\eqref{eq:c_virial} using perturbed potentials, $U_N^\pm = U_N\pm\sigma_N$, $N=2,3$. 
The uncertainty due to errors in the potential is estimated as half of the absolute difference between $C_\varepsilon(T)$ obtained with $U_N^+$ and $U_N^-$. 
The overall uncertainty was then estimated as the sum in quadrature of the uncertainties due to the two-body and three-body potentials. 
The total uncertainty is mainly dominated by the uncertainty from the three-body potential. 
The integrations were performed for all possible triangle configurations with sides up to 13 nm using the adaptive Gauss-Kronrod quadrature of degree 7 and 15. \cite{kronrod1965nodes}

\subsubsection{Results}

\begin{figure}
\centering
\includegraphics[width=1\columnwidth]{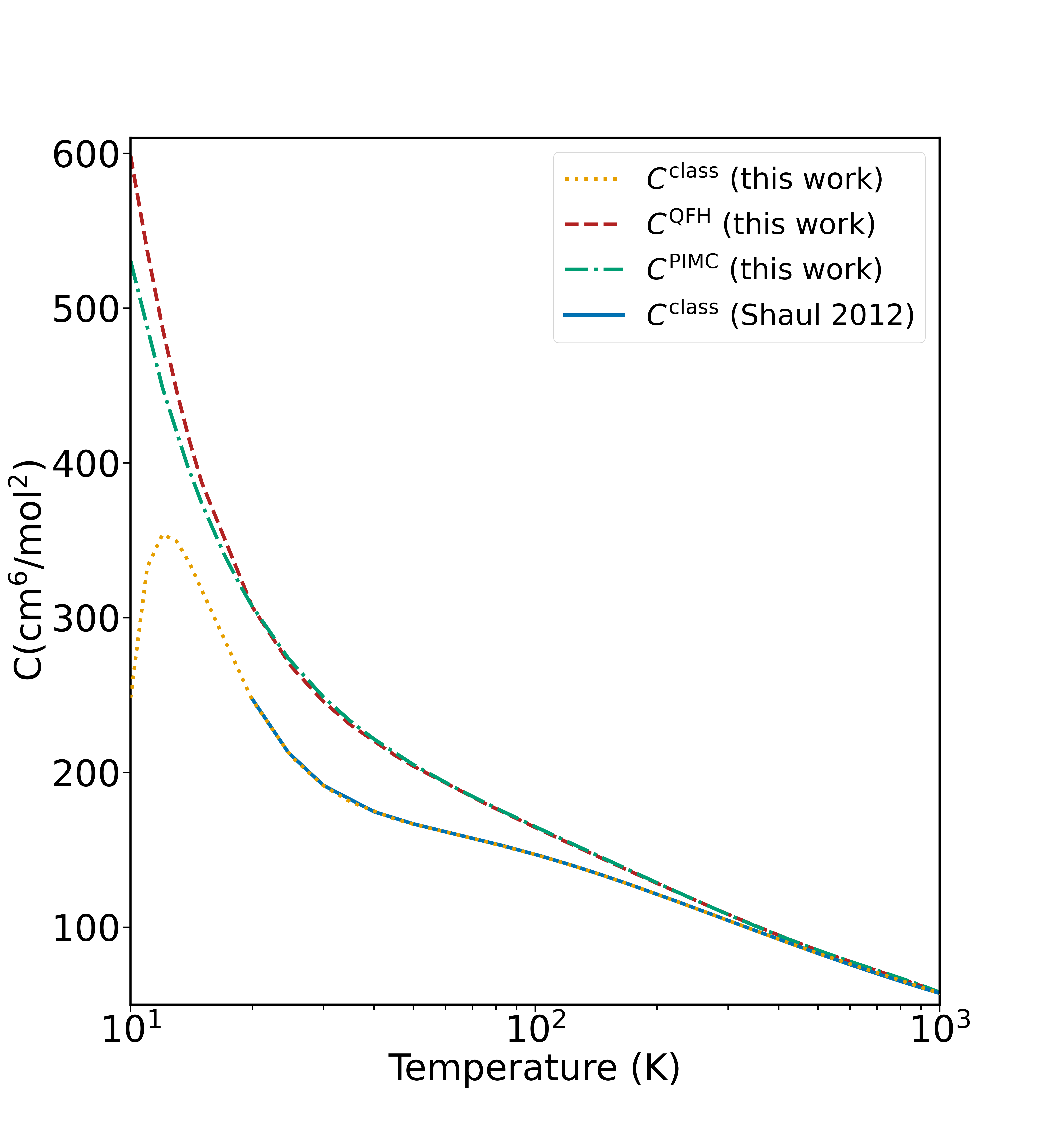}
\caption{
Comparison of the third virial coefficient $C(T)$ calculated using the three-body potential from this work and results from the literature \cite{shaul2012path} that used the CPS09 potential from Ref.~\citenum{Cencek:09}. 
To plot the $C^\mathrm{PIMC}$ curve, results of this work (Table~\ref{tab:C2022}) were interpolated using a function $\sum_{i=1}^6 a_i (T/100)^{b_i}$.}
\label{fig:C_plot}
\end{figure}

\begin{table}
\small
\caption{\ 
Comparison of the third virial coefficients $C(T)$ for $^4$He obtained in this work using the classical and QFH approaches.
The values in parentheses are calculated uncertainties of the rightmost digits.}
\label{tab:C}
\begin{tabular*}{0.48\textwidth}{@{\extracolsep{\fill}}d{4.2}d{3.8}d{3.3}}
\hline
\mcc{Temperature (K)} &
\mcc{$C^\mathrm{class}(T)$(cm${}^6$/mol${}^2$)} &
\mcc{$C^\mathrm{QFH}(T)$(cm${}^6$/mol${}^2$)} \\
&
\mcc{(this work)} &
\mcc{(this work)} \\
\hline
10&	247.90(78)&	598.599\\
11&	332.50(61)&	537.038\\
12&	353.81(49)&	486.976\\
13&	349.37(41)&	446.610\\
14&	335.07(35)&	413.935\\
15&	317.81(30)&	387.256\\
20&	246.72(18)&	306.924\\
25&	210.04(13)&	268.212\\
30&	191.29(10)&	245.654\\
35&	180.815(86)&	230.638\\
45&	169.890(65)&	211.067\\
50&	166.545(58)&	204.015\\
75&	155.447(40)&	180.044\\
100&	146.926(32)&	164.440\\
125&	139.402(27)&	152.638\\
150&	132.683(24)&	143.120\\
175&	126.668(21)&	135.162\\
200&	121.268(20)&	128.347\\
273.15&	108.242(16)&	112.820\\
273.16&	108.241(16)&	112.819\\
300&	104.275(16)&	108.280\\
340&	98.986(15)&	102.329\\
400&	92.176(13)&	94.814\\
500&	83.034(12)&	84.928\\
1000&	57.2801(98)&	57.934\\
1200&	51.3519(93)&	51.842\\
1500&	44.6267(86)&	44.969\\
1700&	41.1131(83)&	41.393\\
2000&	36.8276(80)&	37.042\\
2400&	32.3874(77)&	32.546\\
\hline
\end{tabular*}
\end{table}

\begin{figure}
\centering
\includegraphics[width=1\columnwidth]{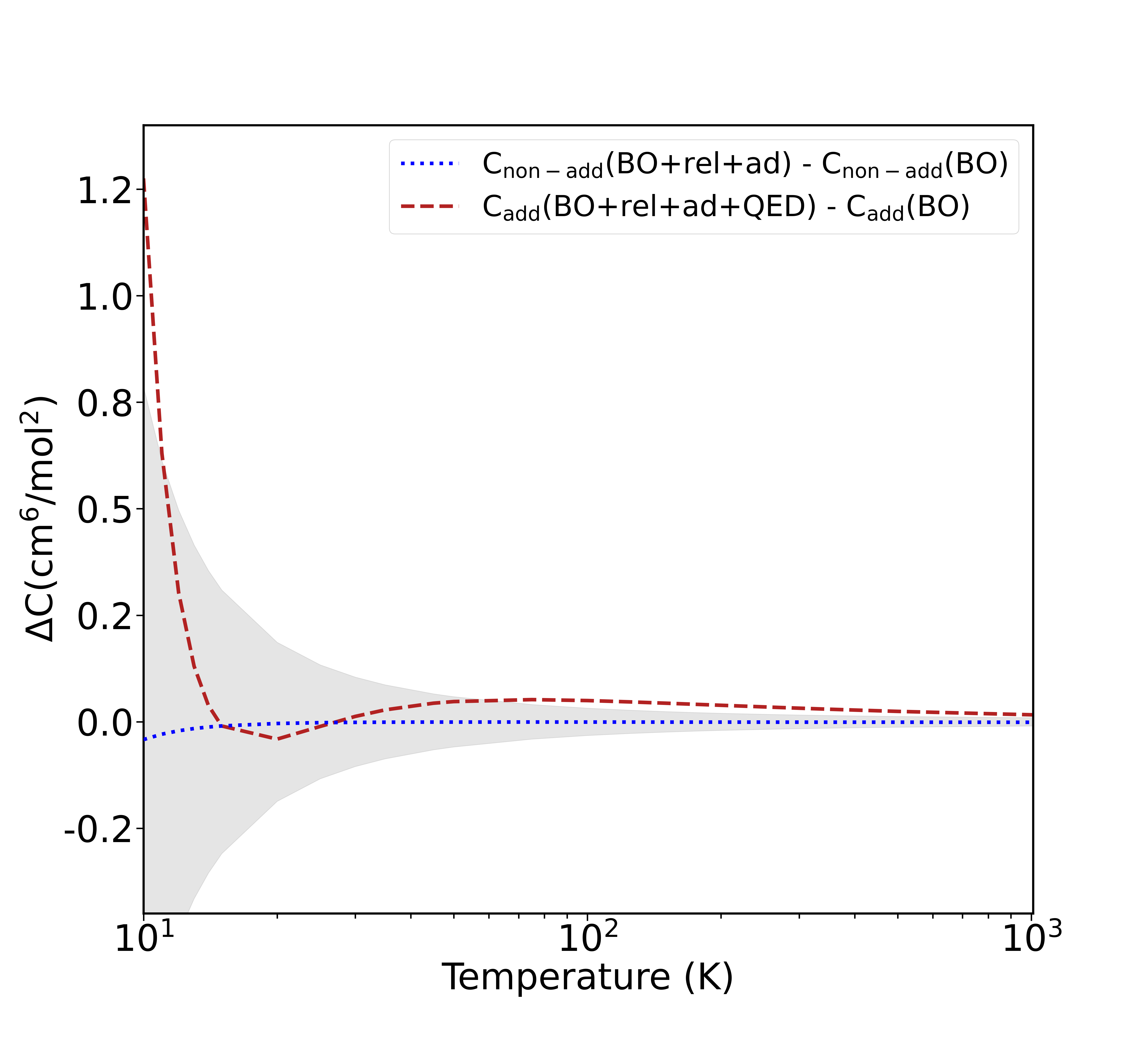}
\caption{
Comparisons of the estimated uncertainty of classical $C(T)$ from this work, represented by the gray area, and contributions of the post-BO corrections to the value of the classical $C(T)$.}
\label{fig:C_plot_sigma}
\end{figure}

In Fig.~\ref{fig:C_plot}, we compare our values of the third virial coefficient with the results taken from the literature. \cite{shaul2012path}
The numerical values together with the calculated uncertainties are presented in Table~\ref{tab:C}.
Our classical results generally agree with the classical results of Shaul \emph{et al.}, \cite{shaul2012path} which were obtained using the CPS09 potential, but are somewhat lower. 
The difference slightly increases with the decreasing temperature but does not exceed 1\%. 
Note that Shaul \emph{et al.} \cite{shaul2012path} did not provide in their work the uncertainties of the classical third virial coefficient due to the propagation of errors of the potentials.
Therefore, we calculated $C^\mathrm{class}(T)$ using the CPS09 potential and its global estimation of uncertainty equal to 2\%. \cite{Cencek:09}
Our present results for the classical third virial coefficient are always within the uncertainty of $C^\mathrm{class}(T)$ calculated this way but our estimated uncertainties are 3-5 times smaller.

Semiclassical QFH approximation is able to fix the behavior of the classical approach, however, the difference between QFH and PIMC values of $C(T)$ are larger than the estimated uncertainty up to 500~K.

The role of the three-body post-BO corrections is small, about 2 orders of magnitude smaller than the estimated uncertainty of $C(T)$. 
By contrast, the importance of the two-body post-BO corrections becomes larger than the estimated uncertainty obtained in this work for temperatures below 15~K and above 75~K, see Fig.~\ref{fig:C_plot_sigma}.

\subsection{Quantum calculations}

We performed quantum path-integral calculations of $C(T)$ at selected temperatures to evaluate the accuracy of the new potential energy surface and include quantum diffraction effects with no uncontrolled approximations.
We used the approach described in Refs.~\citenum{garberoglio2009firstprincip,garberoglio2011improved,Garberoglio2011err,Garberoglio2011a,Garberoglio2011aerr} based on the path-integral formulation of quantum statistical mechanics evaluated using Monte Carlo methods. 
This approach is based on writing the $N$-body partition function, $Z_N = \mathrm{tr}(e^{-\beta H_N})$ -- where $H_N = T_N + V_N$ is the $N$-body Hamiltonian written as a sum of the $N$-body kinetic energy $T_N$ and the $N$-body total potential energy $V_N$ -- using $e^{-\beta H_N} = \left( e^{-\beta H_N / P}\right)^P$ and the Trotter--Suzuki factorization in the limit of large $P$
\begin{equation}\label{eq:primitive}
e^{-\beta H_N/P} \simeq e^{-\beta T_N /P} e^{-\beta V_N/P}.
\end{equation}

Inserting completeness relations between the $P$ factors of $e^{-\beta H_N/P}$ and using the known values of the matrix elements of $e^{-\beta T_N / P}$, \cite{feynman1965pathintegrals, Tuckerman10,garberoglio2009firstprincip} one represents the original partition function of $N$ quantum particles as an equivalent classical partition function of $NP$ classical particles, arranged in $N$ ring polymers of $P$ monomers each, that can be straightforwardly evaluated using Monte Carlo methods. \cite{garberoglio2009firstprincip, Garberoglio2011a}
The mapping is exact in the $P \to \infty$ limit, although in practice one observes convergence for a finite value of $P$ which generally grows larger as the temperature decreases, $P=1$ being the classical limit.

In our path-integral Monte Carlo (PIMC) approach, we calculated virial coefficients as the sum of two contributions: a purely two-body contribution (obtained by assuming that the non-additive part of the three-body potential is zero) and the remainder (which depends on the non-additive three-body potential). 
This procedure, which was first described by Shaul \emph{et al.}, \cite{shaul2012path} optimizes the amount of CPU time required for the calculation. 
In fact, the evaluation of the two-body contribution requires a large number of Monte Carlo samples, which can be efficiently computed using cubic spline interpolation for the pair potential.
The remainder, which performs CPU-time consuming calls to the three-body potential routine, is found to require far fewer Monte Carlo samples to obtain a statistical uncertainty smaller than (or comparable to) that of the two-body contribution.

The propagation of the uncertainty from the potentials was made by using the approach developed by Garberoglio and Harvey, \cite{Garberoglio2021a} which is based on the functional differentiation of the formula for the virial coefficients with respect to the pair or the three-body potential. 
For example, the propagated uncertainty due to the non-additive contribution to the three-body potential, $\delta C^{(3)}$, assuming the classical limit, can be obtained from Eq.~\eqref{eq:c_virial} as
\begin{equation}\label{eq:dC3}
\begin{split}
\delta C^{(3)}(T) 
&= \int \delta U_3 \left| \frac{\delta C}{\delta U_3} \right| ~ dX \\
&= \int \delta U_3 \left| \beta e^{-\beta U_3}\right| ~ dX,
\end{split}
\end{equation}
where we have denoted by $\delta U_3$ the uncertainty of the non-additive three-body potential $U_3$, and we have used the fact that $U_3$ only appears in $Z_3$. 
In Eq.~\eqref{eq:dC3}, $X$ represents all the coordinates used in the classical integration, and we have inserted an absolute value in the functional derivative in order to have a conservative estimate. 
Formulas enabling the propagation of the uncertainty from the pair potential, $\delta C^{(2)}$, can be found in the original publication. \cite{Garberoglio2021a} 
Finally, the uncertainties propagated from the pair and three-body potentials are summed in quadrature. 
Equation~\eqref{eq:dC3} and the equivalent expressions for $\delta C^{(2)}$ have been evaluated using a semiclassical approach where we used the fourth-order Feynman--Hibbs effective pair potential and the unaltered form of the non-additive three-body potential.

As we will discuss extensively later, the new pair and three-body potentials have a much smaller uncertainty than the ones used in our previous works, so much smaller in fact that a brute-force approach to the reduction of the statistical uncertainty would have required very long calculations (recall that the statistical uncertainty in a Monte Carlo calculation falls as the square root of the number of integration points considered), both for $C(T)$ and the acoustic virial coefficient $RT \gamma_\mathrm{a}(T)$.
In this work, we used two novel approaches to reduce the variance of the PIMC calculations.
The first improvement is the use of a higher-order factorization of the high-temperature exponential of Eq.~\eqref{eq:primitive}. 
This enabled the calculation of the two-body contribution to $C(T)$ with a smaller statistical uncertainty for a given amount of computational power. 
Another improvement is related to the calculation of the temperature derivatives needed for the acoustic virial coefficient and is discussed in detail in Sec.~\ref{sec:acoustic}.

\subsubsection{Computational details}

In the case of fully quantum calculations of $C(T)$ we went beyond the so-called primitive approximation for the factorization of the high-temperature density matrix, that is, the approximation that is used in Eq.~\eqref{eq:primitive}, employing the improved propagator derived by Li and Broughton \cite{LB1987} and, independently, by Kono \emph{et al.} \cite{KTL88} based on an initial idea by Takahashi and Imada. \cite{TI84} 
If all the particles present in the system have the same mass $m$, this improved expansion reads
\begin{equation}
e^{-\beta H_N/P} \simeq e^{-\beta T_N /P} e^{-\beta V_N/P} e^{-(\beta/P)^3 O/24},
\end{equation}
with $O$ defined as
\begin{equation}
O = \frac{\hbar^2}{m} \sum_{k=1}^N \left|\nabla_k V_N \right|^2,
\end{equation}
where $\nabla_k$ is the gradient with respect to the position of particle $k$.
Using this approach, we were able to evaluate the two-body contribution to $C(T)$ using a number of replicas $P = \mathrm{nint}(4 + \sqrt{14400~\mathrm{K} / T})$, where $\mathrm{nint}(x)$ denoted the closest integer to the number $x$. 
We used this method only to evaluate the two-body contribution to $C(T)$, which is generally the largest one.
In performing the calculations presented here, our target was to obtain a statistical uncertainty of the order of $1/3$ of the one propagated from the potential. 
To achieve this goal, we needed to average over several independent calculations, each using $10^6$ Monte Carlo calls, using the parallel implementation of the VEGAS algorithm \cite{pvegas} with the integrand averaged over $64$ independent sets of ring polymers. 
The number of independent calculations needed was a decreasing function of the temperature, going from $900$ at $T=10$~K to $30$ at $T=1000$~K.

The calculation of the three-body contribution to $C(T)$ required much less computational resources than the two-body contribution and hence was performed with the same procedure as our previous works, \cite{Garberoglio2011a,garberoglio2011improved} that is using the primitive approximation. 
Due to the higher accuracy of the potential, compared to that used in 2011, we used in this case $P = \mathrm{nint}(14 + 2400~\mathrm{K}/T)$, which is twice as much as what we used previously. 
We reached well converged results using just $4$ independent simulations at each temperature, with $10^5$ Monte Carlo calls and averaging the integrand over $16$ independent ring-polymer
configurations.

\subsubsection{Results and discussion}

\begin{table}[t]
\small
\caption{\ 
Values of the third virial coefficient $C(T)$ for $^4$He obtained in this work, compared with the most recent calculation by the Kofke group. \cite{Schultz2019}
These latter values and their uncertainty were obtained using the analytic fit provided by Gokul \emph{et al.} \cite{Gokul2021} 
Both columns were obtained using the PIMC approach. 
The values in parentheses are calculated uncertainties ($k=2$) of the rightmost digits.}
\label{tab:C2022}
\begin{tabular*}{0.48\textwidth}{@{\extracolsep{\fill}}d{4.2}d{3.7}d{3.6}}
\hline
\mcc{Temperature (K)} & 
\mcc{$C(T)$ (cm${}^6$/mol${}^2$)} & 
\mcc{$C(T)$ (cm${}^6$/mol${}^2$)} \\
    & 
\mcc{(this work)} &
\mcc{(from Gokul \emph{et al.}~\cite{Gokul2021})} \\
    \hline
    10     &   530.6(3)     &  531.3(8)   \\
    15     &   374.10(17)   &  374.5(5)   \\
    20     &   306.85(12)   &  307.1(4)   \\
    30     &   248.65(8     &  248.9(3)   \\
    40     &   221.00(6)    &  222.2(3)   \\
    50     &   205.68(5)    &  205.8(2)   \\
    100    &   164.86(3)    &  164.95(14) \\
    200    &   128.44(2)    &  128.49(9)  \\
    273.16 &   112.87(2)    &  112.91(8)  \\
    300    &   108.32(2)    &  108.36(8)  \\
    400    &   94.840(14)   &  94.87(7)   \\
    500    &   84.944(13)   &  84.97(6)   \\
    1000   &   57.939(10)   &  57.96(5)   \\
    \hline
\end{tabular*}
\end{table}

\begin{figure}
\centering
\includegraphics[width=0.9\linewidth]{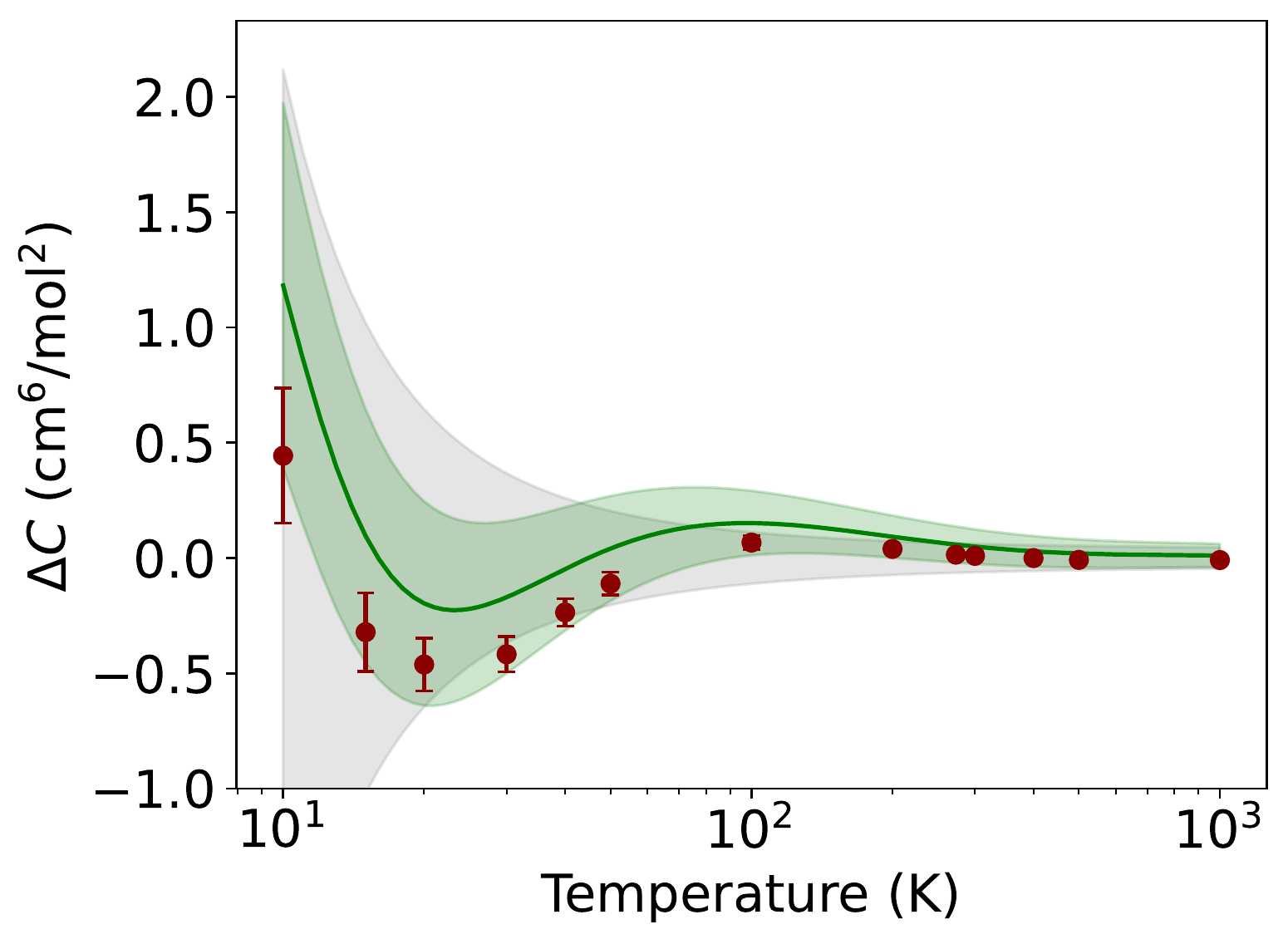}
\caption{
The third virial coefficients of $^4$He reported with respect to the analytic fit developed by Garberoglio \emph{et al.} \cite{garberoglio2011improved}
The gray area in the background is the uncertainty reported by Garberoglio \emph{et al.}, \cite{garberoglio2011improved} the green line are the values obtained from the analytic fit reported by Gokul \emph{et al.} with the surrounding green area representing the uncertainty. \cite{Gokul2021} 
The symbols are the values obtained in the present work. 
All uncertainties are reported at $k=2$ coverage.}
\label{fig:DeltaC}
\end{figure}

Our results for $C(T)$ from PIMC calculations in the temperature range 10--1000~K are shown in Tab.~\ref{tab:C2022}, together with the results of most recent literature calculations, \cite{Schultz2019,Gokul2021} which used the latest pair potential \cite{czachorowski2020second} and the CPS09 three-body potential. \cite{Cencek:09}
The same results are plotted in Fig.~\ref{fig:DeltaC}, where we use as a reference the values from the analytic fit reported by Garberoglio \emph{et al.} \cite{garberoglio2011improved}

In general, our new values for $C(T)$ are very similar to the ones reported in the literature, being compatible between mutual uncertainties.
In particular, our results follow closely the values reported by the Kofke group. \cite{Schultz2019,Gokul2021} 
We notice that the effect of the new three-body potential is a systematic lowering of $C(T)$, at least at the temperatures investigated in the present work. 
Additionally, we notice a significant improvement in the uncertainty budget, which is 3 to 5 times smaller than the previous results. 
Inspection of the contributions to the uncertainties show that generally the propagated uncertainty from the three-body potential is the largest contributor to the overall uncertainty, except close to $T=10$~K, where the propagated uncertainty from the two-body potential becomes of the same order.

When compared to experimental results, the \emph{ab initio} computed virial coefficients have a much smaller uncertainty. 
This was evident even from the first calculations more than 10 years ago, \cite{garberoglio2011improved} and the experimental accuracy has not improved very much in the meantime.
The latest and possibly more accurate experimental determination of $C(T)$ was performed by Gaiser and Fellmuth in 2019, \cite{2019JChPh.150m4303G} who quote the value $C(273.16~\mathrm{K}) = 113.5(1.2)$~cm${}^6$/mol${}^2$ (with an expanded $k=2$ uncertainty).
When compared with our value $112.87(2)$~cm${}^6$/mol${}^2$, the two results are compatible between mutual uncertainties, whereas the uncertainty of the values computed \emph{ab initio} is $\sim 60$ times smaller. 
The improved accuracy of the theoretical values of $C(T)$ presented here resulted in a more accurate primary standard for pressure based on measurement of gas properties. \cite{Gaiser22}

\section{Third acoustic virial coefficient}
\label{sec:acoustic}

The speed of sound $u$ in a low dense gas can be expressed in terms of an expansion of pressure over isotherms \cite{dael1975,gillis1996practical}
\begin{equation}
u^2 = \frac{\gamma_0 RT}{M_\mathrm{m}} 
\left(1 + \frac{ \beta_\mathrm{a}(T)\, p}{RT} + \frac{\gamma_\mathrm{a}(T)\, p^2}{RT} + \frac{\delta_\mathrm{a}(T)\, p^3}{RT} + \cdots \right),
\end{equation}
where $\gamma_0$ is the ratio of the isobaric ($C_\mathrm{p}$) to the isochoric specific ($C_\mathrm{v}$) heats in the limit of zero pressure ($5/3$ for a monoatomic gas), $M_\mathrm{m}$ is a molar mass, and $\beta_\mathrm{a}(T)$, $\gamma_\mathrm{a}(T)$, and $\delta_\mathrm{a}(T)$ are the second, third, and fourth acoustic virial coefficients, respectively.

The pressure and acoustic virial coefficients are connected through a series of equations first presented in Ref.~\citenum{dael1975} and later rederived in Ref.~\citenum{gillis1996practical}. 
With the knowledge of the second and third pressure virial coefficients, the second and third acoustic virial coefficients can be easily obtained from the formulas
\begin{equation}
\beta_\mathrm{a}(T) = 2B(T) + 2(\gamma_0-1)T\frac{dB(T)}{dT} + \frac{(\gamma_0-1)^2}{\gamma_0}T^2\frac{d^2B(T)}{dT^2},
\end{equation}
and
\begin{equation}
RT\gamma_\mathrm{a}(T) = L(T) - \beta_\mathrm{a}(T)B(T),
\end{equation}
with the intermediate $L(T)$ defined as
\begin{align}
\begin{split}
\gamma_0L(T) &= 
   (\gamma_0-1)Q(T)^2 + (2\gamma_0+1)C(T) \\
&+ (\gamma_0^2-1)T\frac{dC(T)}{dT}+\frac{(\gamma_0-1)^2}{2}T^2\frac{d^2C(T)}{dT^2},
\end{split}\\
Q(T) &= B(T) + (2\gamma_0 -1)T\frac{dB(T)}{dT} + (\gamma_0-1)T^2\frac{d^2B(T)}{dT^2}.
\end{align}
Both acoustic virial coefficients depend not only on the pressure virial coefficients but also on their first and second derivatives with respect to temperature. 
In the classical and semiclassical approaches such derivatives are trivial to obtain. 

\subsection{Computational details}

In the case of the quantum approach the straightforward calculation of the temperature derivatives of the path-integral formula for $C(T)$ -- which enter the definition of $RT \gamma_\mathrm{a}(T)$ -- results in expressions that are similar to the so-called thermodynamic estimator of the kinetic energy, as noted in our previous work on this subject. \cite{garberoglio2011improved} 
As is well known, this estimator is characterized by having a very large variance \cite{Tuckerman10} but an alternative formulation, based on the virial theorem, is known to overcome this issue. \cite{Herman82}

The main idea of this formulation is based on recognizing that the matrix elements of the exponential of the kinetic energy operator, which appears in the Trotter--Suzuki splitting of Eq.~\eqref{eq:primitive}, is a homogeneous function of degree two. 
Application of Euler's theorem and successive integrations by parts produce, after lengthy but straightforward derivations, path-integral expressions for the temperature derivatives of $C(T)$, and hence $RT \gamma_\mathrm{a}(T)$, with a significantly reduced variance.

Unfortunately, limitations in the computer resources available at this time prevented us from making the statistical variance smaller than the uncertainty propagated from the potential.
As discussed below, the present calculations have statistical and propagated uncertainties of the same order for $T \geq 273.16$~K, with the situation rapidly deteriorating at lower temperatures. 
We are planning to provide more accurate results, as well as all the details of the derivation of the new approach, in a future paper, which will also cover a more extended temperature range. \cite{C2022}

In the calculations presented here, we used the primitive approximation with the same number of beads as in case of the three-body contribution to $C(T)$. 
Also in this case we found it convenient to compute separately the two-body contribution to $RT \gamma_\mathrm{a}(T)$ and the remainder.
When calculating the two-body contribution, we averaged between 8 and 32 independent configurations, each obtained using $10^6$ Monte Carlo samples with the integrand calculated averaging of $32$ independent realizations of the ring polymers. 
In the case of the three-body contributions, we used a similar approach, but the number of Monte Carlo samples were reduced to $10^5$.

\subsection{Results and discussion}

\begin{table}[t]
\small
\caption{\ 
PIMC values of the third acoustic virial coefficient $RT \gamma_\mathrm{a}(T)$(cm${}^6$/mol${}^2$) obtained in this work, compared with the most recent calculation by Gokul {\em et al.} \cite{Gokul2021}
These latter values and their uncertainty were obtained from the analytic fit provided in the original paper. 
The values in parentheses are calculated uncertainties ($k=2$) of the rightmost digits.}
\label{tab:RTg2022}
  \begin{tabular*}{0.48\textwidth}{@{\extracolsep{\fill}}d{4.2}d{5.2}d{3.6}d{3.6}}
    \hline
\mcc{Temp. (K)} &
\mcc{$RT \gamma_\mathrm{a}(T)$}  &
\mcc{$RT \gamma_\mathrm{a}(T)$} & 
\mcc{$RT \gamma_\mathrm{a}(T)$}  \\
   &
\mcc{(cl. this work)}  & 
\mcc{(PIMC this work)} & 
\mcc{(Gokul \emph{et al.} \cite{Gokul2021})} \\
\hline
  10    & -3054.33  &  808.(10)     &  807.(4)      \\
  15    &   128.75  &  901.(4)      &  901.5(6)    \\
  20    &   551.41  &  794.1(18)   &  795.9(9)    \\
  30    &   561.70  &  590.7(9)    &  591.9(4)    \\
  40    &      -    &  453.5(6)    &  454.1(2)    \\
  50    &   376.50  &  360.2(4)    &  360.53(18)  \\
 100    &   164.94  &  152.01(16)  &  152.22(11)  \\
 200    &    49.01  &   43.09(8)   &   43.24(10)  \\
 273.16 &    19.51  &   15.58(5)   &   15.68(10)  \\
 300    &    12.63  &    9.17(5)   &    9.27(10)  \\
 400    &    -3.81  &   -6.17(4)   &   -6.08(10)  \\
 500    &   -12.64  &  -14.35(4)   &  -14.28(10)  \\
1000    &   -25.87  &  -26.10(3)   &  -26.06(11)  \\
\hline
\end{tabular*}
\end{table}

\begin{figure}
\centering
\includegraphics[width=0.9\linewidth]{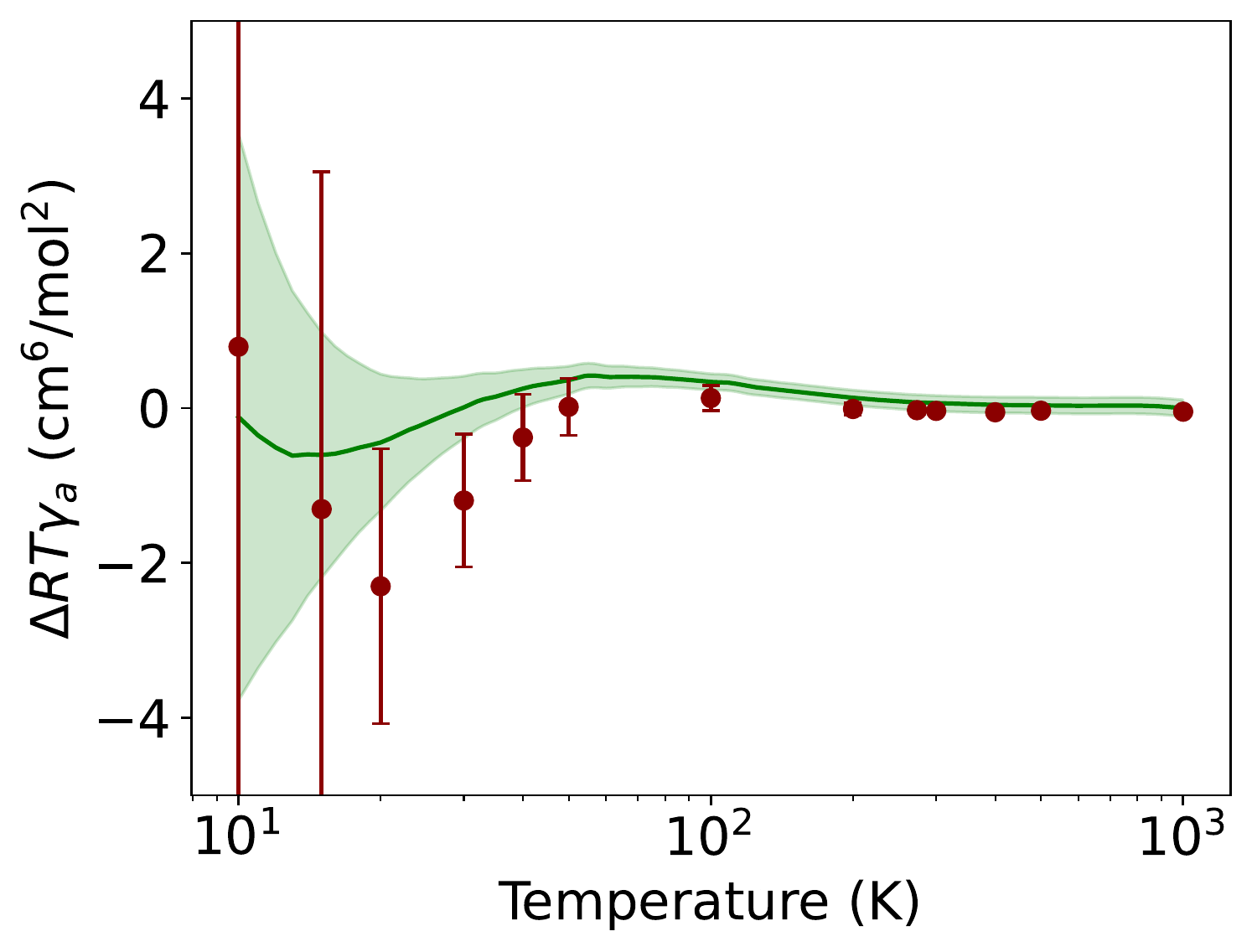}
\caption{The third acoustic virial coefficient $RT \gamma_\mathrm{a}(T)$ of ${}^4$He reported with respect to the analytic fit developed by Garberoglio \emph{et al.} \cite{garberoglio2011improved} 
The green line are the values obtained from the fit reported by Gokul \emph{et al.} \cite{Gokul2021} while the surrounding green area represent their estimated uncertainty. 
The symbols are the values obtained in this work. 
All uncertainties are reported at $k=2$ coverage.}
\label{fig:DeltaRTg}
\end{figure}

The values of the third acoustic virial coefficients $RT \gamma_\mathrm{a}(T)$, obtained in the present work, are shown in Tab.~\ref{tab:RTg2022} for a range of temperatures from 10 K up to 1000 K.
In Fig.~\ref{fig:DeltaRTg} we compare our results with results of the most recent literature calculation. \cite{Gokul2021}
We plot the data from Tab.~\ref{tab:RTg2022} using as a reference the analytic fit developed by Garberoglio \emph{et al.} \cite{garberoglio2011improved} 

The classical and semiclassical results of the third acoustic virial coefficient (the latter ones not shown in Table~\ref{tab:RTg2022}) fail to recover the PIMC results within the uncertainties for all studied temperatures. 
In general, our new PIMC values are compatible with the old ones within mutual uncertainties, although in this case the downward shift already observed in the case of $C(T)$ seems to be amplified.
Regarding the uncertainties, we recall that in this case we did not succeed in making the statistical contribution smaller than the one propagated from the potential, although they are of the same order for $T \geq 273.16$~K. 
Despite this limitation, our uncertainties are smaller, in some cases much smaller, than the ones propagated from the previous three-body potential. 
Given sufficient computational resources, we expect to be able to reduce the uncertainty by a factor $\sqrt{2}$ for $T \geq 273.16$~K (except at $T=1000$~K, where the uncertainty is already dominated by the contribution propagated from the potentials), and up to a factor of $10$ at $T=10$~K.

\section{Conclusions}
\label{sec:conclusions}

We presented \emph{ab initio} calculations of the three-body interaction potential for helium including relativistic corrections and effects due to the coupling of electronic and nuclear motion.
The calculations were performed for 463 points at various levels of theory, and the obtained results were extrapolated to the complete basis set limit to improve their accuracy.
Furthermore, we presented an estimation of uncertainties of our results which is important from the point of view of possible applications in metrology.
Our uncertainty budget contains both the uncertainty due to the basis set incompleteness and due to the missing contribution of higher excitations. 
Our present results were compared to previous calculations of Cencek \emph{et al.} \cite{Cencek:09}
We investigated 50 common configurations and found that the difference between the old and recalculated data was in 60\% of cases larger than the estimated uncertainty of the new results.
Based on the analysis of four points for which an estimation of the uncertainty has been provided in Ref.~\citenum{Cencek:09}, the new uncertainties are 2-4 times smaller.

We constructed global analytic fits of the three-body potentials including correct asymptotic description of the fragmentation processes not only into three isolated atoms but also in the atom-diatom channel.
Our fits of the three-body BO potential and relativistic and adiabatic corrections have the mean absolute relative error with respect to the estimated uncertainties of 0.39$\,\sigma_3^\mathrm{BO}$, 0.29$\,\sigma_3^\mathrm{rel}$, and 0.08$\,\sigma_3^\mathrm{ad}$, respectively.
This correspond to the mean absolute percentage error with respect to the values of the potentials of 0.52\%, 3.78\%, and 2.33\% , respectively.
We compared our fitted potential with the previous best potential from Ref.~\citenum{Cencek:09}.
We found that the new potential is able to accurately fit the dataset from Ref.~\citenum{Cencek:09} while not being specifically constructed using this dataset.
By contrast, the potential from Ref.~\citenum{Cencek:09} fails to properly describe our new dataset mainly due to its insensitivity to changes in linear configurations.

Using classical, semiclassical, and path-integral Monte Carlo methods we calculated the third pressure and acoustic virial coefficients for helium using the potential fit constructed in this work and the best available two-body potentials from Ref.~\citenum{czachorowski2020second}.
Although the difference between the values of classical $C(T)$ and $RT\gamma_\mathrm{a}(T)$ calculated with the potential from Ref.~\citenum{Cencek:09} and with the current potential are relatively small, the estimated uncertainty of the third pressure virial coefficient was reduced by a factor of 3-5 depending on the temperature.
The contribution of the three-body post-BO corrections to the third virial coefficient is smaller than the uncertainty of $C(T)$ due to uncertainties of the BO potential.
We showed that the two-body post-BO corrections become critical for very low or high temperatures as their contribution is almost 2 times larger than the new estimated uncertainty of $C(T)$.

Overall, the PIMC values of $C(T)$ are within the uncertainties of the recent results of Gokul \emph{et al.} \cite{Gokul2021} However, we observed systematical lowering of $C(T)$ when using the new three-body potential in comparison to previous results.
Similarly to the classical results we recovered 3-5 times smaller uncertainties as previously. \cite{Garberoglio2011a,Gokul2021}


\section*{Conflicts of interest}
There are no conflicts to declare.

\section*{Acknowledgements}

We thank Wojciech Cencek and Krzysztof Szalewicz for making available to us the dataset used in Ref.~\citenum{Cencek:09} and Allan H. Harvey for a careful reading of the final manuscript and useful
suggestions on how to improve the presentation. We are also indebted to Michał Lesiuk for useful discussions and encouragement during the preparation of the manuscript.

We acknowledge support from the Real-K project 18SIB02, which has received funding from the EMPIR programme cofinanced by the Participating States and from the European Union’s Horizon 2020 research and innovation programme. 
The support from the National Science Center, Poland, within Project No. 2017/27/B/ST4/02739 is also acknowledged. 
The classical calculations were performed and supported in part by PL-Grid infrastructure and the path-integral calculations were performed on the HPC resources of the University of Trento, which is gratefully acknowledged.

\appendix

\section{Coefficients Z in the asymptotic expansion}
\label{app:Z}

The third-order coefficients $Z_{ijk}$ appearing in the asymptotic expansion of the non-additive three-body BO potential $U_3^\mathrm{BO}$ in the three-atomic fragmentation channel can be calculated from the formula \cite{Bell:70,tang2012long}
\begin{equation}
Z_{ijk} = \frac1\pi \int_0^\infty
\alpha_i(\mathrm{i}\omega)\,\alpha_j(\mathrm{i}\omega)\,\alpha_k(\mathrm{i}\omega)~d\omega,
\end{equation}
where $\alpha_l(\mathrm{i}\omega)$ is the dynamic $2^l$-pole polarizability of an atom at imaginary frequency $\mathrm{i}\omega$. In the sum-over-states representation, the polarizability at imaginary frequency is defined as
\begin{equation}
\alpha_l(\mathrm{i}\omega)=-2\sum_{n\ne0}
\frac{\omega_n\left|\langle n|Q_{l0}|0\rangle\right|^2}{\omega_n^2+\omega^2},
\end{equation}
where the summation goes over all excited states of an atom and $\omega_n=E_0-E_n$ is the deexcitation energy of the state $|n\rangle$. $Q_{l0}$ are the $m=0$ spherical components of the $2^l$-pole moment operator of an atom
\begin{equation}
Q_{lm}=Z\,\delta_{l0}-\sqrt{\frac{4\pi}{2l+1}} \sum_i r_i^l Y_{lm}(\theta_i,\phi_i),
\end{equation}
where $Z$ is the atomic number of the nucleus, the summation is over all electrons whose spherical coordinates are $(r_i,\theta_i,\phi_i)$, and $Y_{lm}(\theta,\phi)$ are the standard, normalized to unity, spherical harmonics.

The leading coefficients in the intra-atomic part of the non-additive three-body post-BO corrections $U_3^\mathrm{Y}$, $\mathrm{Y}\in\{\mathrm{rel},\;\mathrm{ad}\}$, for a system of three identical atoms can be calculated from the formula
\begin{equation}
Z_{111}^{\mathrm{Y},A}=\frac3\pi \int_0^\infty
\alpha_1^2(\mathrm{i}\omega)\,\delta\alpha_1^\mathrm{Y}(\mathrm{i}\omega)~d\omega.
\end{equation}
Corrections to the dynamic dipole polarizability of an atom at imaginary frequency $\mathrm{i}\omega$ are defined by
\begin{equation}
\begin{split}
\delta\alpha_1^\mathrm{Y}(\mathrm{i}\omega)=\,
&-2\sum_{n\ne0}\sum_{n'\ne0}\frac{\omega_{n'}\langle0|G^\mathrm{Y}|n\rangle\langle n|Q_{10}|n'\rangle\langle n'|Q_{10}|0\rangle}{\omega_n(\omega_{n'}^2+\omega^2)} \\
&-2\sum_{n\ne0}\sum_{n'\ne0}\frac{(\omega_n\omega_{n'}-\omega^2)\langle0|Q_{10}|n\rangle\langle n|\overline{G^\mathrm{Y}}|n'\rangle\langle n'|Q_{10}|0\rangle}{(\omega_n^2+\omega^2)(\omega_{n'}^2+\omega^2)} \\
&-2\sum_{n\ne0}\sum_{n'\ne0}\frac{\omega_{n}\langle0|Q_{10}|n\rangle\langle n|Q_{10}|n'\rangle\langle n'|G^\mathrm{Y}|0\rangle}{(\omega_n^2+\omega^2)\omega_{n'}},
\end{split}
\end{equation}
where $\overline{G^\mathrm{Y}}=G^\mathrm{Y}-\langle0|G^\mathrm{Y}|0\rangle$. 
In the case of the relativistic correction, $G^\mathrm{rel}$ is simply equal to the operator $H^\mathrm{rel}$ from Eq.~\eqref{rel:tot} formulated for an atom, but in the case of the adiabatic correction, $G^\mathrm{ad}$ has the following form
\begin{equation}
G^\mathrm{ad}=\frac1{2m}\Big(\sum_i\mathbf{p}_i\Big)^2,
\end{equation}
where the summation is over all electrons in an atom and $m$ is the mass of the atomic nucleus.

The coefficients appearing in the specific part of the asymptotic expansion of the relativistic correction are
\begin{align}
Z_1^{\mathrm{rel},E}
& =-\frac{\alpha^2}\pi\int_0^\infty\omega^2\,\alpha_1^3(\mathrm{i}\omega)~d\omega, \\
Z_{2,3}^{\mathrm{rel},E}
& =-\frac{\alpha^2}\pi\int_0^\infty\omega^2\,\alpha_1^2(\mathrm{i}\omega)\alpha_2(\mathrm{i}\omega)~d\omega, \\
Z_4^{\mathrm{rel},E} 
&=-\frac{\alpha^2}\pi\int_0^\infty\omega^2\,\alpha_1^2(\mathrm{i}\omega)\beta^-_1(\mathrm{i}\omega)~d\omega,
\end{align}
where the generalized polarizability function $\beta^-_1(\mathrm{i}\omega)$ is defined as
\begin{equation}
\beta^-_1(\mathrm{i}\omega)=-2\sum_{n\ne0}\frac{\mathrm{Im}\left[\langle0|Q_{10}|n\rangle\langle n|B^-_{10}|0\rangle\right]}{\omega_n^2+\omega^2},
\end{equation}
with $\mathrm{Im}[x]$ denoting the imaginary part of expression $x$. The operator $B^-_{10}$, in Cartesian representation, has the form
\begin{equation}
B^-_{10}=\sum_i \big(2r_i^2p_{i,z}-z_i(\mathbf{r}_i\cdot\mathbf{p}_i)\big),
\end{equation}
where $\mathbf{r}_i=(x_i,y_i,z_i)$ and $\mathbf{p}_i=(p_{i,x},p_{i,y},p_{i,z})$ are Cartesian components of the position and momentum operators, respectively, of the $i$-th electron.

In Table~\ref{tab:Zcoeff}, we present values of the $Z$ coefficients calculated in this work.
Calculations were performed using one-electron Gaussian basis sets d$X$Z and d$X$Zu described in the main text and FCI description of the wave function of the helium atom.
Our results agree well with the existing literature data for the $Z_{ijk}$ coefficients.

\begin{table}
\small
\caption{\ 
Values of the $Z$ coefficients calculated in this work compared with existing literature data. 
Values from Ref.~\citenum{tang2012long} were divided by 3 due to the difference in the definition of the $Z_{ijk}$ coefficients.
}
\label{tab:Zcoeff}
\begin{tabular*}{0.48\textwidth}{@{\extracolsep{\fill}}cd{2.10}d{2.12}}
\hline
coefficient & \mcc{this work} & \mcc{Ref.~\citenum{tang2012long}} \\
\hline
$Z_{111}$ &  0.49316 & 0.493186202143 \\
$Z_{112}$ &  0.9242  & 0.924267920597 \\
$Z_{122}$ &  1.7394  & 1.73943799007  \\
$Z_{113}$ &  4.125   & 4.126262337997 \\
$Z_{222}$ &  3.288   & 3.2884931971   \\
$Z_{114}$ & 34.131   & \\
$Z_{123}$ &  7.7796  & \\
\\[-1ex]
$Z_{111}^{\mathrm{rel},A}$ & -9.7891 \times 10^{-5} & \\
$Z_{111}^{\mathrm{ad},A}$  &  6.044  \times 10^{-4} & \\
\\[-1ex]
$Z_1^{\mathrm{rel},E}$     & -1.0048 \times 10^{-5} & \\
$Z_{2,3}^{\mathrm{rel},E}$ & -2.1192 \times 10^{-5} & \\
$Z_4^{\mathrm{rel},E}$     & -2.7985 \times 10^{-5} & \\ 
\hline
\end{tabular*}
\end{table}

\section{Angular factors W in the asymptotic expansion}
\label{app:W}

Asymptotic expansions of the non-additive three-body BO potential published thus far involved angular factors expressed in terms of cosines of linear combinations of internal angles in the triangle formed by atoms, $\theta_1$, $\theta_2$, and $\theta_3$. \cite{Bell:70,Doran:71,tang2012long,Lotrich:97} 
This formulation is computationally inefficient as it requires frequent calls to routines calculating trigonometric functions.
In this work, we present an alternative form of the angular factors. 
Our new equations are expressed exclusively in terms of polynomials in cosines of separate angles.
The cosines can be computed once, and then used in the construction of all terms. 
The equivalence of both formulations can be verified using standard trigonometric relations and the identity
\begin{equation}
\cos^2\theta_1+\cos^2\theta_2+\cos^2\theta_3=1-2\cos\theta_1\cos\theta_2\cos\theta_3,
\end{equation}
which holds since $\theta_1+\theta_2+\theta_3=\pi$ for internal angles of a planar triangle.
Furthermore, we present angular factors in the asymptotic expansion of the fourth-order terms in the three-body BO potential, and leading terms in the asymptotic expansion of the post-BO corrections. 
In all presented formulas, we use the notation $c_i=\cos\theta_i$, $i=1,2,3$, and $C=c_1c_2c_3$.

The third-order BO angular factors $W_{ijk}$ are:
\begin{align}
W_{111} 
& = 3 + 9C,
\\
W_{112}
& = \frac32 \big(c_3 (9 + 15C - 5c_3^2) - 3c_1c_2\big),
\\
W_{122}
& = \frac{15}4 \big(c_1 (-2 + 20C + 5c_1^2) + 5c_2c_3 (2 + 7C)\big),
\\
W_{113} 
& = \frac52 \big(-3 - 9C + 7c_3^2 (3 + 3C - 2c_3^2)\big),
\\
W_{222} 
& = \frac{15}8 \big(18 + 110C + 245C^2
  - 35 (c_1^2c_2^2 + c_2^2c_3^2 + c_3^2c_1^2)\big),
\\
W_{114} 
& = \frac{15}8 \Big(c_3 \big(-27 - 42C + 7 c_3^2 (14 + 9C - 9c_3^2)\big) + 3c_1c_2\Big),
\\
\begin{split}
W_{123} 
& = \frac{15}4 \Big(c_2 \big(-6 - 35C - 5c_1^2 + 7c_3^2 (4 + 21C + 3c_2^2)\big) \\
& + c_3c_1 (19 + 56C - 21c_3^2)\Big).
\end{split}
\end{align}

The fourth-order BO angular factors $W_1,\dots,W_4$ are:
\begin{align}
W_1
& = \frac1{30\sqrt{15}} \big(c_1 (-9 - 12C + 4c_1^2 + 9c_3^2) + 6c_2c_3\big),
\\
W_2
& = \frac1{90} \big(c_3 (3 - 30C - 10 c_3^2) - 3c_1c_2 (4 + 15C)\big),
\\
W_3
& = \frac1{75} (-7 - 6C + 15 c_3^2),
\\
W_4
& = \frac2{25} (-1 - 3C).
\end{align}

The angular factors $W_1^{\mathrm{rel},E}$, $W_2^{\mathrm{rel},E}$, and $W_3^{\mathrm{rel},E}$ appearing in the specific part of the three-atomic asymptotic expansion of the relativistic correction are:
\begin{align}
W_1^{\mathrm{rel},E} 
& = \frac12 (5 + 3C - 12c_3^2),
\\
W_2^{\mathrm{rel},E} 
& = \frac34 \big(c_3 (23 + 5C - 35c_3^2) - c_1c_2\big),
\\
W_3^{\mathrm{rel},E} 
& = \frac34 \big(c_3 (3 + 3C - c_3^2 - 6c_1^2) - 3c_1c_2\big).
\end{align}

\bibliography{c_density_paper.bib}

\end{document}